\documentstyle[a4p,12pt,epsf,rotating,cite]{article}
\def\mathrm{\rm}

\def\Journal#1#2#3#4{{#1}{\bf #2} (#3) #4}

\def\NIMA{{\em Nucl. Instr. and Meth.} A}
\def\NPB{{\em Nucl. Phys.} B}
\def\PLB{{\em Phys. Lett.}  B}

\def\PRD{{\em Phys. Rev.} D}
\def\ZPC{{\em Z. Phys.} C}
\def\EPC{{\em Eur. Phys. J.} C}
\newcommand{\bc}{\begin{center}}
\newcommand{\ec}{\end{center}}
\newcommand{\bi}{\begin{itemize}}
\newcommand{\ei}{\end{itemize}}

\def\Zz{\mbox{$\mathrm{Z}^0$}}

\newcommand{\vnbr}[2] { \begin{array}{c} #1\\#2 \end{array} }
\newcommand{\textvl}[2] {\begin{array}{c} \mathrm{#1}\\ \mathrm{#2} \end{array}}

\newcommand{\upfs}{\Upsilon(\mathrm{4S})}
\newcommand{\Lb}{\mbox{$\Lambda_{\mathrm b}$}}

\newcommand{\fD}{\mbox{${f(\mathrm c \to D^{\pm},~D^0)}$}}
\newcommand{\fDs}{\mbox{${f(\mathrm c \to D_s)}$}}
\newcommand{\fLc}{\mbox{${f(\mathrm c \to \Lambda_c)}$}}
\newcommand{\Ds}{\mbox{${\mathrm D_s}$}}
\newcommand{\Dzero}{\mbox{${\mathrm D^0}$}}
\newcommand{\Dplus}{\mbox{${\mathrm D^{\pm}}$}}
\newcommand{\Bs}{\mbox{${\mathrm B_s}$}}
\newcommand{\Bo}{\mbox{${\mathrm B^0}$}}
\newcommand{\Bpl}{\mbox{${\mathrm B^{\pm}}$}}

\newcommand{\gbb}{\mbox{${\mathrm g \to b {\bar b}}$}}
\newcommand{\gcc}{\mbox{${\mathrm g \to c {\bar c}}$}}

\newcommand{\Bell} {\mbox{${\mathrm BR}({\mathrm B}\rightarrow {\mathrm X}\ell$)}}

\newcommand{\brl} {\mbox{${\mathrm BR}({\mathrm b}\rightarrow \ell{\mathrm X}$)}}

\newcommand{\bre} {\mbox{${\mathrm BR}({\mathrm b}\rightarrow e {\mathrm X}$)}}
\newcommand{\bru} {\mbox{${\mathrm BR}({\mathrm b}\rightarrow \mu {\mathrm X}$)}}
\newcommand{\brcl} {\mbox{${\mathrm BR}({\mathrm b}\to {\mathrm c} \to \ell {\mathrm X}$)}}
\newcommand{\brce} {\mbox{${\mathrm BR}({\mathrm b}\to {\mathrm c} \to e {\mathrm X}$)}}
\newcommand{\brcu} {\mbox{${\mathrm BR}({\mathrm b}\to {\mathrm c} \to \mu{\mathrm X}$)}}

\newcommand{\bbbar}{\mbox{$\mathrm {b \bar b}$}} 
\newcommand{\ccbar}{\mbox{$\mathrm {c \bar c}$}}

\newcommand{\fbl}{\mbox{$f({\mathrm b} \to \ell)$}}
\newcommand{\fbe}{\mbox{$f({\mathrm b} \to {\mathrm e})$}}
\newcommand{\fbu}{\mbox{$f({\mathrm b} \to \mu)$}}

\newcommand{\fbcl}{\mbox{$f({\mathrm b} \to {\mathrm c} \to \ell)$}}
\newcommand{\fbce}{\mbox{$f({\mathrm b} \to {\mathrm c} \to {\mathrm e})$}}
\newcommand{\fbcu}{\mbox{$f({\mathrm b} \to {\mathrm c} \to \mu)$}}

\newcommand{\btaul}{\mbox{${\mathrm b} \to \tau \to \ell$}}

\newcommand{\jpsi}{\mbox{${\mathrm J}/\psi$}}
\newcommand{\jpsil}{\mbox{${\mathrm J}/\psi \to \ell^+\ell^-$}}
\newcommand{\bjpsil}{\mbox{${\mathrm b} \to {\mathrm J}/\psi \to \ell^+\ell^-$}}

\newcommand{\cl}{\mbox{${\mathrm c} \to \ell$}}

\newcommand{\btoul}{\mbox{${\mathrm b} \to X_u \ell \nu$}}
\newcommand{\nnbl}{\mbox{$\mathrm{NN}_{{\mathrm b} \ell}$}}
\newcommand{\nnbcl}{\mbox{$\mathrm{NN}_{{\mathrm b}{\mathrm c} \ell}$}}
\newcommand{\Bl}{\mbox{${\mathrm BR}({\mathrm B}\rightarrow {\mathrm X}\ell$)}}
\newcommand{\bl}{\mbox{${\mathrm b} \to \ell$}}
\newcommand{\be}{\mbox{${\mathrm b} \to {\mathrm e}$}}
\newcommand{\bu}{\mbox{${\mathrm b} \to \mu$}}
\newcommand{\bcl}{\mbox{${\mathrm b} \to {\mathrm c} \to \ell$}}
\newcommand{\bcbarl}{\mbox{${\mathrm b} \to {\mathrm {\bar c}} \to \ell$}}

\newcommand{\peterson}{\epsilon_b}

\newcommand{\bce}{\mbox{${\mathrm b} \to {\mathrm c} \to {\mathrm e}$}}
\newcommand{\bcu}{\mbox{${\mathrm b} \to {\mathrm c} \to \mu$}}
\newcommand{\Ks}{\mbox{$K_s^0$}}

\newcommand{\ebb}{\eta_{\mathrm b}}
\newcommand{\ecc}{\eta_{\mathrm c}}
\newcommand{\euds}{\eta_{\mathrm uds}}
\newcommand{\effl}{\mbox{{$\epsilon_{\mathrm b \to \ell}$}}}
\newcommand{\effe}{\mbox{{$\epsilon_{\mathrm b \to e}$}}}
\newcommand{\effu}{\mbox{{$\epsilon_{\mathrm b \to \mu}$}}}

\newcommand{\effcl}{\mbox{{$\epsilon_{\mathrm b \to \mathrm{c} \to \ell}$}}}
\newcommand{\effce}{\mbox{{$\epsilon_{\mathrm b \to \mathrm{c} \to e}$}}}
\newcommand{\effcu}{\mbox{{$\epsilon_{\mathrm b \to \mathrm{c} \to \mu}$}}}
\newcommand{\rb}{\mbox{$R_{\mathrm b}$}}
\newcommand{\rc}{\mbox{$R_{\mathrm c}$}}

\newcommand{\xe}{\mbox{$\langle x_E \rangle$}}
\newcommand{\xec}{\mbox{$\langle x_E \rangle_{\mathrm c}$}}
\newcommand{\nc}{\mbox{$\langle n_c \rangle$}}

\newcommand{\nbl}{\mbox{${N_{\mathrm b \to \ell}}$}}

\newcommand{\nb}{\mbox{$N_{\mathrm b}$}}
\newcommand{\Nt}{\mbox{$N_{\mathrm t}$}}
\newcommand{\Ntt}{\mbox{$N_{\mathrm tt}$}}
\newcommand{\nbtag}{\mbox{$N_{\mathrm b-tags}$}}
\newcommand{\nol}{\mbox{$N_{\ell}$}}

\newcommand{\pb}{\mbox{$P_{\mathrm b}$}}
\newcommand{\pf}{\mbox{$p_F$}}
\newcommand{\mc}{\mbox{$m_c$}}
\newcommand{\dds}{\mbox{${\mathrm D^{**}}$}}
\newcommand{\msp}{\mbox{${\mathrm m_{sp}}$}}

\newcommand{\Zzero}{\mbox{$\mathrm{ Z}^0$}} 
\newcommand{\dedx} {{\mathrm d} E/{\mathrm d}x}

\newcommand{\ts}{\thinspace}
\newcommand{\etal}{{\it et al.}}

\newcommand{\mevocc} {\mathrm \thinspace MeV/c^2}
\newcommand{\mevoc} {\mathrm \thinspace MeV/c}
\newcommand{\gevocc} {\mathrm \thinspace GeV/c^2}
\newcommand{\gevoc} {\mathrm \thinspace GeV/c}
\newcommand{\gev} {\mathrm \thinspace GeV}
\newcommand{\pc} {\thinspace\%}

\renewcommand{\arraystretch} {1.2}

\begin{document}
\pagenumbering{roman}
\begin{titlepage}
\begin{center}{\large   EUROPEAN LABORATORY FOR PARTICLE PHYSICS}\end{center}\bigskip
 \begin{flushright}
%   OPAL PR279\\
  CERN-EP/99-078\\
  $8^{\mathrm{th}}$ June 1999
 \end{flushright}
\bigskip\bigskip
\begin{center}{\huge\bf  Measurements of inclusive \\
     semileptonic branching fractions\\
     of b hadrons in $\boldmath \Zz \unboldmath$ decays. 
}\end{center}\bigskip\bigskip
\begin{center}{\LARGE The OPAL Collaboration
}\end{center}\bigskip\bigskip
\bigskip\begin{center}{\large  Abstract}\end{center}
\noindent {A measurement of inclusive semileptonic branching fractions of
     b hadrons produced in Z$^0$ decays is presented. An enriched Z$^0 \to
     \bbbar$ sample is
     obtained with a lifetime flavour-tagging technique. The leptonic
     events are then selected from
     this sample, and classified according to their 
     origin, which is 
     determined by comparing the distribution of several
     kinematic variables using artificial neural network techniques.
     Using 3.6 million multihadronic events collected with the OPAL
     detector at energies near the Z$^0$ resonance, the values
\begin{eqnarray*}
\brl  & = & (10.83\pm 0.10{\mathrm~(stat.)}\pm 0.20{\mathrm~(syst.)}^{~+0.20}_{~-0.13}{\mathrm~(model)})\pc \\
\brcl & = & (8.40\pm 0.16{\mathrm~(stat.)}\pm 0.21{\mathrm~(syst.)}^{~+0.33}_{~-0.29}{\mathrm~(model)})\pc
\end{eqnarray*}
    are  measured, where b denotes all weakly decaying b hadrons and
    $\ell$ represents either $e$ or $\mu$. The second error includes
    all experimental systematic uncertainties whereas the last error
    is due to uncertainties in modelling of the 
    lepton momentum spectrum in semileptonic decays and b quark fragmentation. 
    The average fraction of the beam energy carried by the 
weakly decaying b hadron, $\xe$, is measured to be
\begin{eqnarray*}
\xe = 0.709 \pm 0.003 {\mathrm~(stat.)}\pm 0.003 {\mathrm~(syst.)} \pm 0.013 {\mathrm~(model)} 
\end{eqnarray*}
where the modelling error is dominated by the choice of b
fragmentation model. 
The agreement between data and
various semileptonic decay models and fragmentation functions
is also investigated.
}
\bigskip\bigskip\bigskip\bigskip
\bigskip\bigskip
\begin{center}{\large
(Submitted to European Physics Journal {\bf C})
}\end{center}
\end{titlepage}
\begin{center}{\Large        The OPAL Collaboration
}\end{center}\bigskip
\begin{center}{
%begin authorlist PLEASE DO NOT DELETE THIS COMMENT
G.\thinspace Abbiendi$^{  2}$,
K.\thinspace Ackerstaff$^{  8}$,
G.\thinspace Alexander$^{ 23}$,
J.\thinspace Allison$^{ 16}$,
N.\thinspace Altekamp$^{  5}$,
K.J.\thinspace Anderson$^{  9}$,
S.\thinspace Anderson$^{ 12}$,
S.\thinspace Arcelli$^{ 17}$,
S.\thinspace Asai$^{ 24}$,
S.F.\thinspace Ashby$^{  1}$
% ...
%end authorlist PLEASE DO NOT DELETE THIS COMMENT
}\end{center}\bigskip
\bigskip
%begin institutes
$^{  1}$School of Physics and Astronomy, University of Birmingham,
Birmingham B15 2TT, UK
\newline
$^{  2}$Dipartimento di Fisica dell' Universit\`a di Bologna and INFN,
I-40126 Bologna, Italy
\newline
% ...
$^{  5}$Cavendish Laboratory, Cambridge CB3 0HE, UK
\newline
% ...
$^{  8}$CERN, European Organisation for Particle Physics,
CH-1211 Geneva 23, Switzerland
\newline
$^{  9}$Enrico Fermi Institute and Department of Physics,
University of Chicago, Chicago IL 60637, USA
\newline
% ...
$^{ 12}$Indiana University, Department of Physics,
Swain Hall West 117, Bloomington IN 47405, USA
\newline
% ...
$^{ 16}$Department of Physics, Schuster Laboratory, The University,
Manchester M13 9PL, UK
\newline
$^{ 17}$Department of Physics, University of Maryland,
College Park, MD 20742, USA
\newline
% ...
$^{ 23}$Department of Physics and Astronomy, Tel Aviv University,
Tel Aviv 69978, Israel
\newline
$^{ 24}$International Centre for Elementary Particle Physics and
Department of Physics, University of Tokyo, Tokyo 113-0033, and
Kobe University, Kobe 657-8501, Japan
\newline
% ...
\newline
%end notes
\newpage
\pagenumbering{arabic}
%%%%%%%%%%%%%%%%%%%%%%%%%%%%%%%
\section{Introduction}
%%%%%%%%%%%%%%%%%%%%%%%%%%%%%%%
\label{sec:intro}
\leavevmode\indent
Measurements of the semileptonic branching fraction of b hadrons are
important in testing our understanding of the dynamics of heavy quark
physics and are also important inputs for other b physics analyses. These
measurements can also be used to extract the CKM matrix element V$_{cb}$.
Recent QCD calculations which include higher-order
perturbative corrections \cite{ball,neubert} have lowered the
predicted value of the semileptonic branching ratio for B mesons,
${\mathrm{BR}}_{SL}^B$, and now adequately reproduce the experimental
results \cite{pdg98}. These calculations
also predict a value for $\nc$, the average number of
charmed hadrons produced per B meson decay,
which is consistent with experimental data \cite{neubert}.

While theoretical calculations are now in better agreement with
experiment, the measurements of ${\mathrm BR}_{SL}$
obtained at the $\upfs$ and Z$^0$ resonances 
have slightly disagreed for some time. 
The semileptonic branching fraction for B mesons has been measured 
at the $\upfs$ resonance
to be ${\mathrm BR}_{SL}^{\mathrm B} = (10.45 \pm 0.21)\pc$ \cite{pdg98},
whereas a combination of all results obtained at the Z$^0$ resonance
gives ${\mathrm BR}_{SL}^{\mathrm b} = (10.99 \pm
0.23)\pc$ \cite{pdg98}, where
the superscript b indicates that the high energy data correspond to
a mixture of $\Bpl$, $\Bo$, $\Bs$ and b baryons,  as opposed to $\Bpl$
and $\Bo$ only at the $\upfs$ resonance. 
Assuming the semileptonic width, $\Gamma_{\mathrm sl}$, to be the same
for all b hadrons, as suggested by the result of~\cite{sllb}, and
given that the semileptonic branching ratio is related to the
lifetimes, $\tau$, by ${\mathrm BR}_{SL} = \Gamma_{\mathrm
sl}/\Gamma_{\mathrm total} = \tau \Gamma_{\mathrm sl}$, one obtains
\begin{equation}
{\mathrm BR}_{SL}^{\mathrm B}  = {\tau_{\mathrm B} \over \tau_{\mathrm
b}} \cdot {\mathrm BR}_{SL}^{\mathrm b} \label{eq:br1}.
\end{equation}

To remove the difference between the $\upfs$ and Z$^0$
results would 
require that ${\tau_{\mathrm B}/\tau_{\mathrm b}}$ be
$0.948$, whereas current lifetime measurements yield
% show that the b baryon
% lifetime is less than the B meson lifetime, and hence that 
${\tau_{\mathrm B}/\tau_{\mathrm b}}$ greater than
one~\cite{pdg98}. After applying the 
correction factor ${\tau_{\mathrm B}/\tau_{\mathrm b}}$,
there is a difference of about two standard deviations
between the $\upfs$ and Z$^0$ results for the branching fractions.

This paper describes the measurement of the semileptonic branching
fraction for all b hadrons produced at or near the $\Zzero$ resonance
using identified electrons and muons in an enriched $\Zzero \rightarrow
\bbbar$ sample.
The measurement differs from previously published
measurements \cite{OPAL,ALEPH,DELPHI,L3} 
in the use of a method designed specifically to achieve a precise
determination of the semileptonic branching ratios rather than
extracting them from a multi-dimensional fit for several
electroweak parameters. 
The dependence on the semileptonic decay models has been
substantially reduced and the correlation with the $\rb$ measurement 
eliminated. 
The analysis also allows the determination of $\brcl$ and $\xe$,
the mean fraction of beam energy carried by the weakly decaying b
hadron, both of which are important inputs needed for other heavy
flavour measurements, such as those described in \cite{bogus}.
The agreement between data and
various semileptonic decay models and fragmentation functions
is investigated in the Appendix.
% 
%%%%%%%%%%%%%%%%%%%%%%%%%%%%%%%%%%%%%%%%%%%%%%%%%%%%%%%%%%%%%%%%%%%%%%%
\section{The OPAL detector, data and Monte Carlo samples}
%%%%%%%%%%%%%%%%%%%%%%%%%%%%%%%%%%%%%%%%%%%%%%%%%%%%%%%%%%%%%%%%%%%%%%%%
%
The OPAL detector is described in
reference \cite{detector}. The central tracking system
is composed of a silicon microvertex detector,
a precision vertex drift chamber, and a large-volume jet
chamber surrounded by a set of drift chambers that measure 
the $z$-coordinate.\footnote{The
coordinate system is defined such that the $z$-axis follows the
electron beam direction and the $x$-axis points towards the centre 
of the LEP ring.  The polar
angle~$\theta$~is defined relative to the $+z$-axis, and the
azimuthal angle~$\phi$~is defined relative to the $+x$-axis.}
Charged particles are identified by their specific energy
loss, $\dedx$, in the jet chamber gas.  Further information on the
performance of the tracking and $\dedx$ measurements can be found in
reference \cite{dedx}.
These detectors are located inside a solenoid providing a magnetic
field of 0.435 T.
Outside the solenoid are a time-of-flight
scintillator array and a lead-glass electromagnetic calorimeter with a
presampler.
Including the endcap electromagnetic calorimeters,
the lead-glass blocks cover the range $| \cos \theta | < 0.98$.
The next layer is the hadron calorimeter, consisting of the
instrumented return yoke of the magnet. Finally, the detector is
covered by several layers of muon chambers. In total, at least seven,
and in most regions eight, absorption lengths 
of material lie between the interaction point and the muon detectors.
This material is sufficient to absorb low-momentum muons produced at
the vertex, 
but most muons with momenta above 
$2 \gevoc$ reach the muon detectors.
The muon chambers are constructed as two different detector
subsystems in the barrel and endcap; together,
they cover $93\pc$ of the full solid angle.

This analysis uses events recorded at centre-of-mass energies within
$3 \gev$ of the Z$^0$ peak during the
$1992-1995$ running period when the silicon microvertex detector was fully
operational. A total of 7.15 million multihadronic Monte Carlo events, 
and 4.88 and 1.93 million simulated $\bbbar$ and $\ccbar$
hadronic decays are also analysed.
All samples were produced with the JETSET~7.4
Monte Carlo generator \cite{jetset}, with the fragmentation
function of Peterson \etal\ \cite{peterson} for heavy quarks.
The ACCMM model \cite{ACCMM} tuned to the CLEO data \cite{cltune}
is used to describe the lepton momentum spectrum in $\bl$ and
$\bcl$ decays, as described in \cite{EWWG-NIM}.
The Monte Carlo parameters were tuned to describe the OPAL data \cite{opalmc}. 
All simulated events were passed through the full OPAL detector simulation
package \cite{gopal}.
% 
%%%%%%%%%%%%%%%%%%%%%%%%%
\section{Event selection and analysis method}
%%%%%%%%%%%%%%%%%%%%%%%%%
\leavevmode\indent
Standard hadronic event selection \cite{opalmh} and detector
performance requirements are applied to select a sample of 3.35 million events where
the primary vertex can be reconstructed.
Each event is divided into two hemispheres by the plane perpendicular
to the thrust axis and containing the interaction point. 
The thrust value of the event is required to be greater than 0.8 to
suppress contributions from events containing more than two jets, 
for example from hard gluon radiation or
events with gluon splitting into a heavy quark pair. The
polar angle of the thrust axis, $\theta_{\mathrm th}$, must satisfy
$\vert\cos\theta_{\mathrm th}\vert<0.75$, to ensure that the event is
contained within the central barrel region. 
A total of 2.15 million events satisfy these requirements. 

Lifetime tagging techniques are used to suppress the contributions from
non-$\bbbar$ events. 
Hemispheres are tagged as containing b hadrons
(``b-tagged'') using a neural network algorithm \cite{rb98}.
A cut is applied to the network output, selecting a
sample of b-hemispheres with a purity of $92\pc$ and an
efficiency of $30\pc$.
The b purity for the hemisphere b-tag is extracted directly from the
data, as detailed in the next section.

A search for lepton candidates is made in the hemisphere opposite a
b-tagged hemisphere in events containing one or two such hemispheres.
By using leptons found in the hemisphere opposite the b-tagged
hemisphere rather than in the b-tagged hemisphere itself,
a sample is obtained which does not bias 
the relative fraction of the different b hadron species.
In addition, this method avoids introducing significant correlations
between b flavour tagging and lepton selection. 

Jets are formed from charged tracks and electromagnetic
energy clusters unassociated with tracks using a cone algorithm
\cite{cone}, with a minimum energy of $5 \gev$ and a cone radius of
550 mrad. Jet shape and momentum variables are used in two 
neural networks, $\nnbl$ and $\nnbcl$, trained respectively to
distinguish direct decays, $\bl$, and cascade decays, $\bcl$, 
from all other lepton sources, collectively termed as backgrounds.
Separate neural networks are trained for electrons and muons.
Details of the training of these neural networks are given in Section
\ref{sec:anntrain}. The distributions of the neural network outputs
are compared for the data and the Monte
Carlo to determine the fractions of events
from $\bl$ and $\bcl$ decays. Contributions from direct
$\cl$ decays are suppressed by the b-tagging requirement.

To determine the fraction of leptons coming from
$\bl$ and $\bcl$ decays, a binned log-likelihood fit is performed
which uses the shapes of the distributions of the neural network
output variables. The fit also yields $\xe$.
The number of $\bl$ and $\bcl$ decays is obtained by multiplying
the fitted fractions by the number of selected leptons,
corrected by the lepton detection efficiencies, derived from Monte Carlo. 
Dividing the numbers of $\bl$ and $\bcl$
candidates by the number of b-flavoured hemispheres 
yields the branching ratios $\brl$ and $\brcl$.
The number of b-flavoured hemispheres is extracted from the data using
a double tagging technique.
%           ---------------------
\subsection{Purity of the sample of b-tagged hemispheres}
\label{bpurity}
%           ---------------------
The purity $\pb$ of the b-tagged sample of hemispheres is extracted from
the data using a double-tagging technique, thereby minimising systematic uncertainties.
The number of hemispheres $\Nt$ passing the
b-flavour tagging criteria described above is counted together with
the number of events $\Ntt$ where both hemispheres are b-tagged. 
With the b-tagging efficiencies for the b, c and light flavours given by $\ebb$,
$\ecc$ and $\euds$, one can write
\begin{equation}
\Nt=2 N_{\mathrm MH} [ \rb \ebb + \rc \ecc + (1 -\rb -\rc) \euds)], \label{eq:nt}
\end{equation}
\begin{equation}
\Ntt=N_{\mathrm MH} [ C_{\mathrm b} \rb \ebb^2 + C_{\mathrm c} \rc
\ecc^2 + C_{\mathrm uds} (1 -\rb -\rc) \euds^2], \label{eq:ntt}
\end{equation}
where $N_{\mathrm MH}$ represents the number of events that passed
the multihadronic selection and fiducial cuts. $\rb$ and $\rc$ are the
fractions of multihadronic $\Zzero$ events decaying into $\bbbar$ and
$\ccbar$ pairs, respectively. The hemisphere correlation coefficients
$C_{\mathrm q}$, where q represents any primary quark flavour, are
given by the ratio $C_{\mathrm q} = \eta_{\mathrm qq} /
\eta_{\mathrm q}^2$, where $\eta_{\mathrm q}$ is the
efficiency for tagging a hemisphere containing flavour q and
$\eta_{\mathrm qq}$ is the efficiency for
tagging both hemispheres. Equations \ref{eq:nt} and \ref{eq:ntt} can
be re-expressed in terms of the purity $\pb$ instead of $\ebb$ using
the definition  $\pb = 2 N_{\mathrm MH} \rb \ebb /\Nt$, and then used
to directly measure the b purity.
Whilst the b purity can be determined from the direct solution
of either Equation \ref{eq:nt} or Equation \ref{eq:ntt} separately, 
the value of $\pb$ is extracted by maximising
the log-likelihood of both equations simultaneously to obtain the
maximum statistical sensitivity. 

A total of $N_{\mathrm MH} = 2 \thinspace 150 \thinspace 423$ events
passed the multihadronic event selection. These events contained 
$\Nt = 303 \thinspace 366$ b-tagged hemispheres, of which 
$\Ntt = 45 \thinspace 351$ have also the other hemisphere
b-tagged. The world average values of $\rb$ and $\rc$ \cite{pdg98} are
used as inputs.  To extract $\pb$ from Equations
\ref{eq:nt} and \ref{eq:ntt}, the charm and light-flavour
efficiencies $\ecc$ and $\euds$ and
the correlation for b events, $C_{\mathrm b}$,
are taken from Monte Carlo, while $C_{\mathrm c}$ and $C_{\mathrm uds}$, 
which have negligible impact on the b purity determination, are taken to be
unity.
From the fit to the data, the purity of all b-tagged
hemispheres is measured to be $(91.901 \pm 0.016)\pc$, where the error is
statistical. 

Extensive studies have 
been presented in a previous OPAL analysis on the systematic differences
between data and Monte Carlo for $C_{\mathrm b}$, $\ecc$ and $\euds$
\cite{rb98}.  
Similar studies were conducted for this analysis.
This was performed separately for the data collected in 1992 ($19\pc$ of the
total data sample) where 
only $r-\phi$ information was available from the silicon microvertex detector,
and for the larger part of the data, where $z$ information was also
available. The corresponding weighted averages for 
the combined sample are used as input parameters to the purity
fit. 
The systematic uncertainties on $\pb$ are summarised in
Table \ref{systbpur}. The purity is
\begin{equation}
\pb = (91.90 \pm 0.02 {\mathrm~(stat.)} \pm 0.45{\mathrm~(syst.)})\pc \nonumber
\end{equation}
for a b-tagging efficiency $\ebb$ of around $30\pc$.
\begin{table}
\begin{center}
\begin{tabular}{|c|c|c|} \hline
Input parameter & value                     & $\delta\pb$  \\ \hline\hline
  \rc   &$0.177 \pm 0.008$                                                   &$\pm 0.19\pc $ \\
  \rb   &$0.2169 \pm 0.0012$                                                 &$\pm 0.01\pc $ \\  
  $\ecc$  &$0.0209 \pm 0.0002{\mathrm~(MC~stat.)} \pm 0.0014 {\mathrm~(syst.)}$&$\pm 0.34\pc $ \\ 
  $\euds$ &$0.0034 \pm 0.0000{\mathrm~(MC~stat.)} \pm 0.0003 {\mathrm~(syst.)}$&$\pm 0.22\pc $ \\ 
${\mathrm C_b}$ &$1.0493 \pm 0.0052{\mathrm~(MC~stat.)} \pm 0.0052 {\mathrm~(syst.)}$&$ \pm 0.01\pc $ \\ \hline
total           &                          & $ \pm 0.45\pc $ \\ \hline 
\end{tabular}
\parbox{15cm}{\caption {\sl
Contributions to the systematic uncertainty $\delta\pb$ on
the b purity \pb.
For the input parameters taken from the Monte Carlo, the
statistical and systematic uncertainties are shown separately. 
An error of $0.0000$ indicates a value less than $0.00005$.
\label{systbpur}}}
\end{center}
\end{table}
%
%           ---------------------
\subsection{Electron identification}
\label{sec:eid}
%           ---------------------
\leavevmode\indent
High-momentum electrons are searched for
in the hemisphere opposite to a b-tagged hemisphere.
Electrons are identified using an artificial neural network \cite{rb98}.
The six inputs used by this neural network are: the momentum and polar
angle of the track; the ratio $E/p$ of the electromagnetic energy and
track momentum;
the number of electromagnetic calorimeter blocks contributing to the
energy measurement; the normalised ionization energy loss
$\dedx|_{\mathrm norm}$ and
its error. The normalised $\dedx$ value is defined as 
$\dedx|_{\mathrm norm}=(\dedx-\dedx|_0)/\sigma_0)$, where
$\dedx|_0$ is the ionization energy loss expected for an electron and
$\sigma_0$ the error.

Candidate tracks must have a minimum of 40 jet chamber hits usable for the
determination of the energy loss, out of a possible 159 hits.
Electrons are required to have momentum greater than $2 \gevoc$.
The momentum cut is applied to reduce the fraction of fake electrons
and to restrict the analysis to the region where the input variables
used for this neural network are properly modelled.
Electrons from photon conversion and electrons from Dalitz decays are
rejected as in \cite{rb98}. 

In the data sample, $301 \thinspace 303$ b-tagged hemispheres were
selected when the sub-detectors required for electron identification
were fully operational and $29 \thinspace 516$ electrons were found in
the opposite hemispheres.
Monte Carlo events are used to determine the electron identification efficiency
after b-tagging for $\be$ and $\bce$ decays. 
These are measured to be
$\effe  =  0.5662 \pm 0.0012$ and  
$\effce  =  0.3306 \pm 0.0012$, 
where the errors come from Monte Carlo statistics.
The effect of the momentum cut at $2 \gevoc$ is
taken into account.
%           ---------------------
\subsection{Muon identification}
\label{sec:muid}
%           ---------------------
\leavevmode\indent
All hemispheres opposite a b-tagged hemisphere are searched for
muon candidates.
Muons are identified by associating central detector tracks with track
segments in the muon detectors in two orthogonal planes \cite{muid}. 
The muon candidates
are also required to have momenta greater than $2 \gevoc$.
In total, $44 \thinspace 832$ muons are found in the hemispheres 
opposite $302 \thinspace 577$ b-tagged hemispheres selected
when the muon chambers were functional.
The efficiencies for identifying muons in $\bu$ and $\bcu$ decays with
momentum greater than $2 \gevoc$ are
$\effu   =  0.6794 \pm 0.0011$ and
$\effcu  =  0.4277 \pm 0.0013$, where the errors come 
from Monte Carlo statistics.
% 
%%%%%%%%%%%%%%%%%%%%%%%%%%%%%%%%%%%%%%%%%%%%%%%%%%%%%%
\section{Composition of the lepton sample}
%%%%%%%%%%%%%%%%%%%%%%%%%%%%%%%%%%%%%%%%%%%%%%%%%%%%%%
%
The main contributions to the sample after selecting a lepton in a
b-tagged event come from  direct decays, $\bl$, and cascade decays,
$\bcl$. All remaining sources are collectively referred to as
background.

Fake muons form the largest source of background in the muon sample.
These fake muons are mostly due to light mesons passing through the
hadronic calorimeter without showering. Fake electrons are less common
and consist mostly of mis-identified pions. These fake leptons tend to
have less transverse momentum than leptons from either $\bl$ or
$\bcl$ decays. 
True leptons in $\bbbar$ events not originating from the
semileptonic decay of a b or c quark, for example electrons from
photon conversions, are grouped together as non-prompt leptons.
The decays $\bcbarl$, where the $\bar{\mathrm c}$ comes
from the virtual $W$ boson decay, form an important background to
$\bcl$ decays. The lepton coming from either of the two cascade decays tends
to be produced with less transverse momentum with respect to the jet
axis than a lepton coming from a direct $\bl$ decay. 
The selected lepton samples also contain small contributions from 
$\btaul$ and $\mathrm{b} \rightarrow \jpsi \rightarrow \ell^+\ell^-$
decays.
A smaller contribution to the backgrounds comes from leptons from charm
and light flavour decays. Due to the high b purity of the selected
data sample, this source is greatly suppressed. Finally, a very small
contribution comes from gluon splitting into \ccbar\ and \bbbar.
The sources of background found in the Monte Carlo sample are shown in
Table \ref{bgnd}.

\begin{table}
\begin{center}
\begin{tabular}{|c|c|c|} \hline  
Monte Carlo lepton candidates & $104 \thinspace 653~e$ & $173 \thinspace 735~\mu$ \\ \hline\hline
\bl                           & $54.7 \pc$     & $43.0\pc$  \\
\bcl                          & $27.1 \pc$     & $23.0\pc$  \\
fake leptons                  &  $3.3 \pc$     & $18.5\pc$  \\ 
non-prompt leptons            &  $5.3 \pc$     & $7.3\pc$ \\
\bcbarl                       &  $3.5 \pc$     & $3.1\pc$  \\
\btaul                        &  $2.5 \pc$     & $1.8\pc$  \\ 
\bjpsil                       &  $0.9 \pc$     & $0.7\pc$  \\ 
primary \ccbar~events         &  $2.4 \pc$     & $2.2\pc$  \\ 
primary uds  events           &  $0.2 \pc$     & $0.2\pc$  \\ 
\gcc                          &  $0.2 \pc$     & $0.1\pc$  \\ 
\gbb                          &  $< 0.001\pc$  & $< 0.001\pc$  \\ \hline 
\end{tabular}
\parbox{15cm}{\caption {\sl
Composition of the Monte Carlo sample of electron and muon candidates
opposite a b-tagged hemisphere showing the contributions from
$\bl$, $\bcl$ and background.
\label{bgnd}}}
\end{center}
\end{table}

Instead of attempting to reject these backgrounds, a fit
for the fractions of $\bl$ and $\bcl$ decays in the sample is
performed using the two-dimensional distribution of the
output variables of two neural networks:
The first neural network, $\nnbl$, is trained to separate $\bl$ events from
all other events while the second, $\nnbcl$, is trained to separate 
$\bcl$ events from all other events. 
Each of the neural networks 
is trained separately for electron and muon samples
since the background is different in the two.
% 
%%%%%%%%%%%%%%%%%%%%%%%%%%%%%%%%%%%%%%%%%%%%%%%%%%%%%%
\subsection{Neural network training}
\begin{figure}[htbp]\centering
 \begin{center}
   \mbox{
    \epsfxsize=\textwidth
    \epsffile{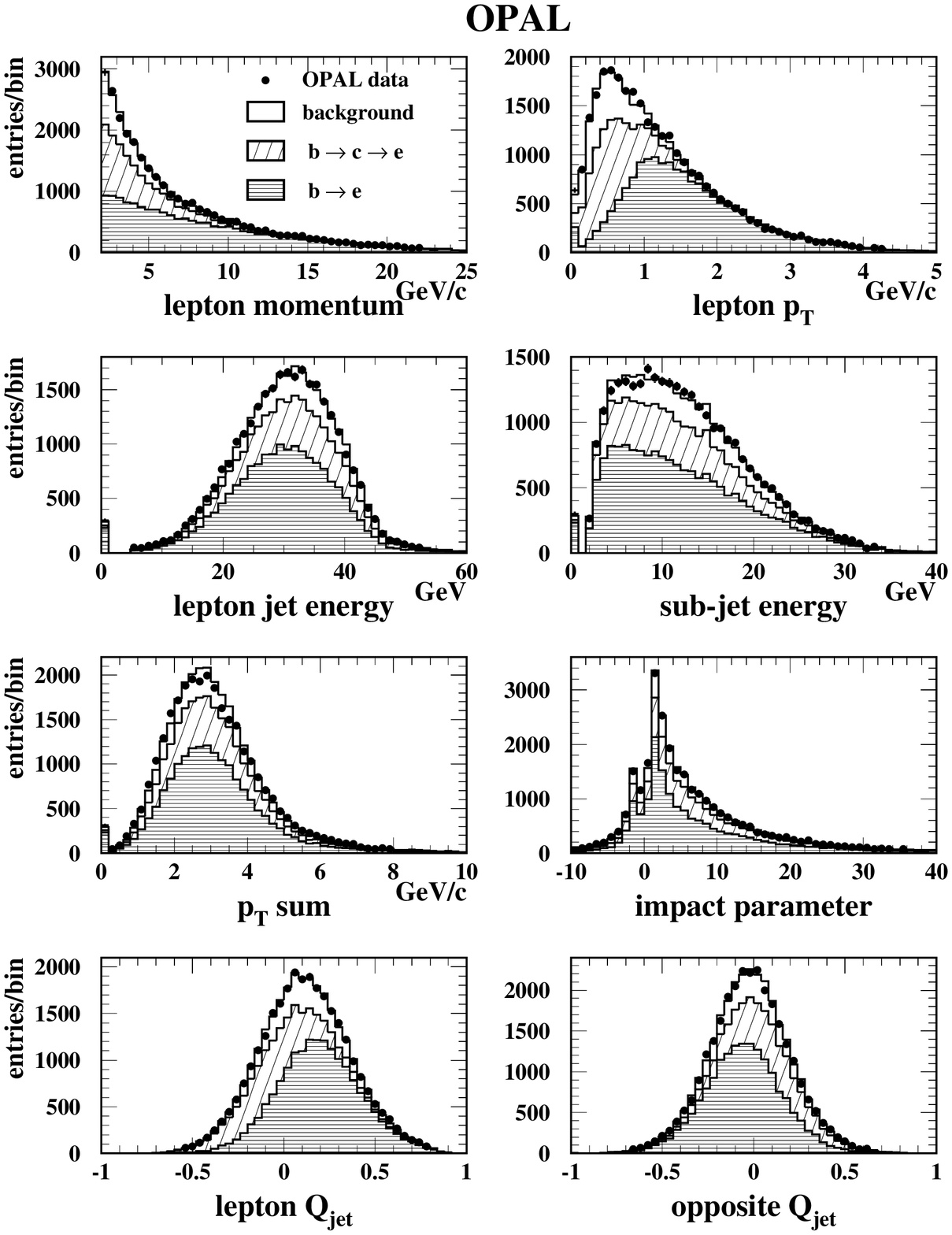}
   }
 \parbox{15cm}{\caption {\sl
The eight input variables used by the neural networks $\nnbl$ and
$\nnbcl$ to separate $\bl$ and $\bcl$ decays. The contributions from
the three classes of Monte Carlo events; $\bl$, $\bcl$ and
backgrounds,  are superimposed and compared to
the data after normalisation and rescaling using the fractions
$\fbl$ and $\fbcl$ derived in Section \ref{sec:fitresults}. The results
are shown here for electrons only.
\label{nninpute}}}
\end{center}
\end{figure}

%%%%%%%%%%%%%%%%%%%%%%%%%%%%%%%%%%%%%%%%%%%%%%%%%%%%%%
\label{sec:anntrain}
%           ---------------------
Jet and lepton kinematic variables in b-tagged events are used as
input variables to train the neural networks $\nnbl$ and $\nnbcl$ to
select direct and cascade leptons coming from b decays. 
By combining the information from several variables, 
more discrimination power is obtained than by using momentum
information alone. 
About $90 \thinspace 000$ muon and $70 \thinspace 000$ electron
candidate tracks reconstructed in
Monte Carlo events opposite a b-tagged hemisphere, were used to train 
$\nnbl$ and $\nnbcl$.
Eight kinematic variables were used, which are shown in Figure \ref{nninpute}.
\bi
\item lepton momentum;
\item lepton $p_T$: the lepton transverse momentum calculated with respect to the
  nearest jet axis, excluding the lepton candidate itself;
\item lepton jet energy: the energy of the jet containing the lepton;
\item sub-jet energy: the energy of the sub-jet (defined below) containing the lepton;
\item $p_T$ sum: the scalar sum of transverse momenta of all charged tracks in the
lepton jet;
\item impact parameter: the impact parameter significance of the candidate lepton track with
respect to the primary vertex;
\item lepton $Q_{jet}$; the lepton charge multiplied by the jet
  charge (defined below) of the jet containing
the lepton, including the lepton candidate track;
\item opposite $Q_{jet}$: the lepton charge multiplied by the jet charge of the most
energetic jet in the hemisphere opposite the lepton.
\ei

In $\bl$ decays, the lepton momentum spectrum reflects the hard
fragmentation of the primary b hadron and is thus
efficient at separating these leptons from other sources. The boost
along the b jet direction results in a higher lepton momentum for
$\bl$ than for other decays. Similarly,
the decaying b hadron imparts more $p_T$ to the lepton 
in $\bl$ decays than in $\bcl$ decays. 

The total energy of the jet has sensitivity to leptons from direct
decays since b
jets are expected to have lower visible energy in semileptonic decays
due to the emission of an energetic neutrino. 

The smaller mass of c hadrons relative to b hadrons forces
the non-leptonic decay products from a charm decay to follow the
lepton direction more closely. The neutrino in a charm decay also
carries less energy on average than the neutrino in a primary $\bl$
decay. These differences mean that the energy deposited by
neutral and charged particles in the vicinity of the lepton candidate,
the lepton sub-jet energy, will be lower in $\bl$ decays than in
$\bcl$ decays. The lepton jet is therefore divided into 
two sub-jets \cite{subjet},
where the initial sub-jet seeds are the lepton track and the other
tracks in the jet. Each track and
unassociated electromagnetic cluster is then reassigned iteratively
until each one is closer in angle to its assigned sub-jet axis than to the
other. No track or cluster is added to the sub-jet containing the
lepton beyond an invariant mass upper limit of $2.5 \gevocc$. The
``sub-jet energy'' used for the neural network input refers to the
sub-jet including the lepton.

The scalar sum of $p_T$ for the jet characterises
both the angular width and the multiplicity of the jet, both of which are
known to differ significantly between b and charm and light-quark 
jets \cite{OPAL-sumpt}.

The lepton impact parameter significance is the distance of closest
approach of the track to the primary vertex divided by the uncertainty on
this distance. Larger impact parameter significances are expected for
leptons from decays such as $\bl$ and $\bcl$ decays, than
for tracks from the primary vertex such as fragmentation tracks.

The final two variables consist of the reconstructed lepton charge
multiplied by the jet charge for the jet associated 
with the lepton and for the most energetic jet in the b-tagged
hemisphere. The jet charge is defined as
\begin{equation}
Q_{jet} = \frac{\displaystyle{\sum_{i}  Q_{i} \cdot
{p_{i}}^{0.5}} } {\displaystyle\sum_{i} {p_{i}}^{0.5}}
\end{equation}
where $Q_{i}$ is the track charge, $p_{i}$ is the track momentum
and the sum runs over the charged tracks in the jet including
the lepton candidate. 
Leptons from $\bl$ decays have the same charge as the
weakly decaying b quark and thus the lepton $Q_{jet}$ variable
shows a positive correlation between the lepton charge and associated
jet charge. Leptons from $\bcl$ decays have
opposite charge to the decaying b quark and hence show a negative
correlation with the lepton jet charge. Leptons from $\bcbarl$ decays
have a positive correlation with the lepton jet charge. 
In the absence of $\mathrm{B}^{0}- \overline{\mathrm{B}^{0}}$ mixing,
the correlations between the lepton charge in one hemisphere and the
jet charge in the opposite hemisphere, embodied in the opposite
$Q_{jet}$ variable, are opposite to those of the jet
associated with the lepton.

The distributions of the eight input variables used by the neural networks
are shown for electrons in Figure \ref{nninpute}.
The same good agreement between data and Monte Carlo simulation
is found for muons for all input variables.
Combining the information from
these variables using neural networks allows not only to increase the
separation power 
but also to include the correlations between the input variables.

\begin{figure}[htbp]\centering
 \begin{center}
   \mbox{
    \epsfxsize=\textwidth
    \epsffile{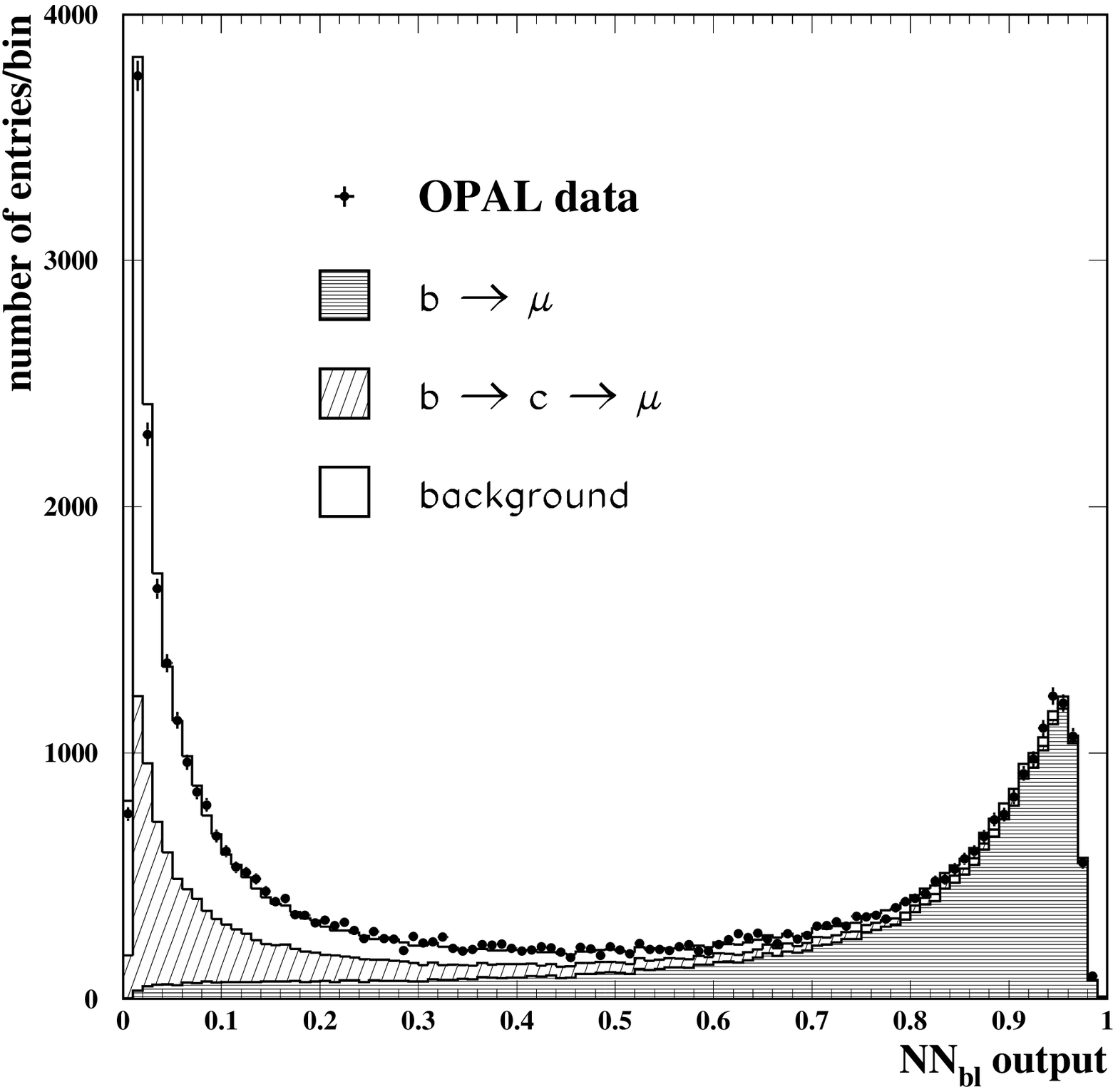}
   }
 \parbox{15cm}{\caption {\sl
Distribution of the output variable for the neural network $\nnbl$ trained
specifically to distinguish $\bl$ events from all other types of
events. The three categories $\bl$, $\bcl$, and background are
described in the text.   The results are shown here for muons only.
\label{fittedbu}}}
\end{center}
\vspace*{-0.8cm}
\end{figure}

\begin{figure}[htbp]\centering
 \begin{center}
   \mbox{
    \epsfxsize=\textwidth
    \epsffile{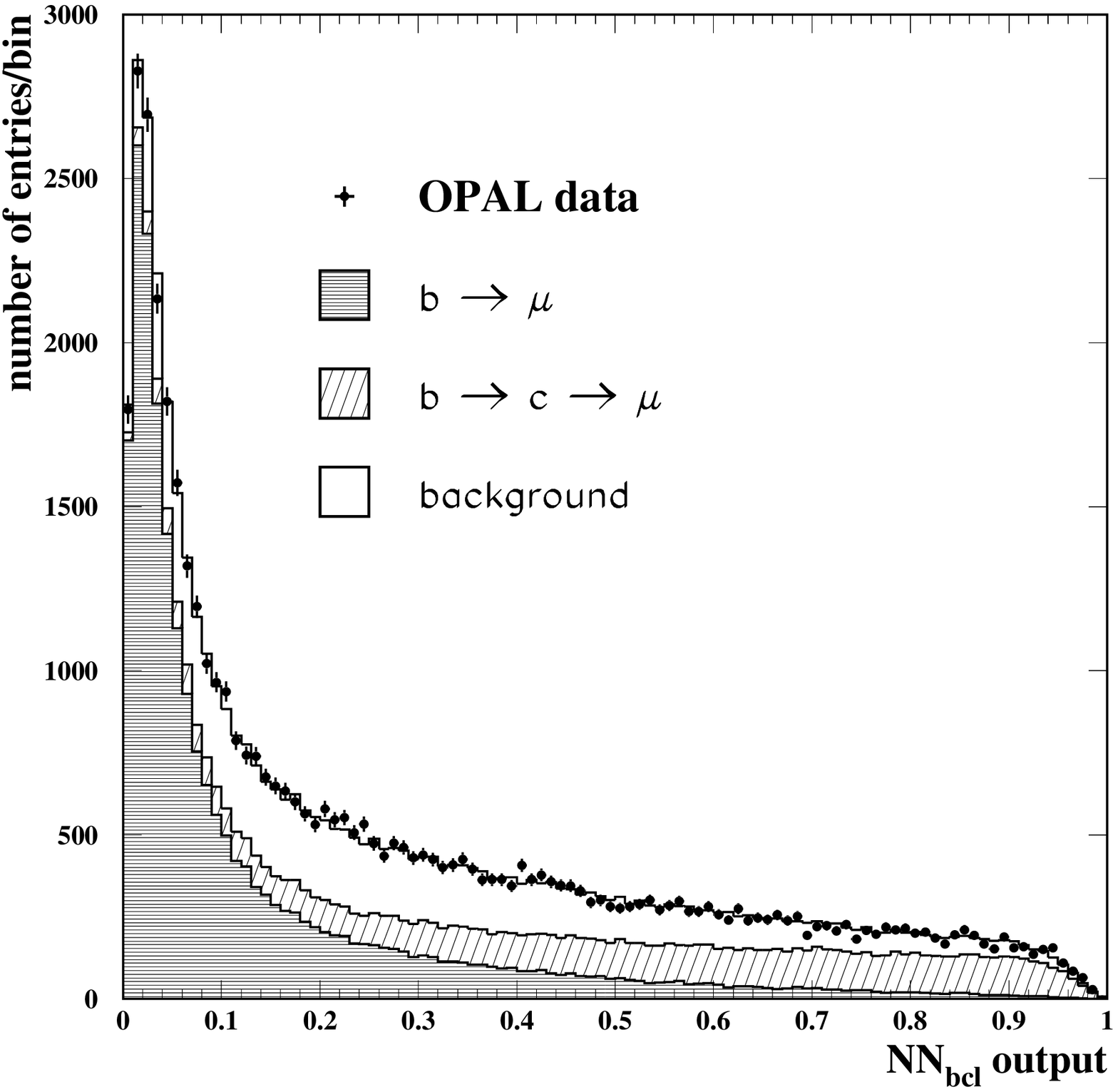}
   }
 \parbox{15cm}{\caption {\sl
Distribution of the output variable for the neural network $\nnbcl$
trained specifically to distinguish $\bcl$ events from all other types
of events. The three categories $\bl$, $\bcl$, and background are
described in the text.   The results are shown here for muons only.
\label{fittedbcu}}}
\end{center}
\vspace*{-0.8cm}
\end{figure}

The distribution of the neural network outputs from $\nnbl$
and $\nnbcl$ for muons are shown in Figures \ref{fittedbu} and \ref{fittedbcu}.
The results are shown % for muons only
for the three categories of Monte Carlo events: direct b decays $\bl$,
cascade decays $\bcl$, and the background. As an indicator of the
goodness-of-fit, the $\chi^2$/bin is $1.00$ 
in Figure \ref{fittedbu} and $0.88$ in Figure \ref{fittedbcu}
and is calculated using the statistical, systematic and
modelling errors (see Section \ref{sec:fitsyst}). 


As can be seen from Figure \ref{fittedbu},
the neural network $\nnbl$ gives good separation of
$\bl$ decays from all other types of decays found
in the b-tagged sample; there is much less separation power between
$\bcl$ events and the background. The second neural
network, $\nnbcl$, gives further discrimination between $\bcl$ events and all
other events as can be seen in Figure \ref{fittedbcu}. 
% 
%%%%%%%%%%%%%%%%%%%%%%%%%%%%%%%%%%%%%%%%%%%%%%%%%%%%%%
\section{Calculation of the semileptonic branching ratios}
%%%%%%%%%%%%%%%%%%%%%%%%%%%%%%%%%%%%%%%%%%%%%%%%%%%%%%
%           --------------------------------------------
\leavevmode\indent
The semileptonic branching fraction is given by
\begin{equation}
\brl= {\nbl \over \nb} = {\nol \cdot \fbl \over \effl} \cdot {1 \over
\nbtag \cdot \pb}
\label{eqn:BR}
\end{equation}
where $\nbl$ is the number of hemispheres containing a semileptonic b
decay and $\nb$ the total number of true b hemispheres.
The fraction of $\bl$ events determined by the fit, $\fbl$, times the
number of lepton candidates, $\nol$, yields the number of $\bl$ decays
in the selected lepton sample. The total number of b events that decayed
semileptonically, $\nbl$, is then obtained by correcting this number
to account for the lepton detection efficiency, $\effl$. The 
number of true b decays in the b-tagged sample, $\nb$, is obtained
from the total number of
b-tagged hemispheres, $\nbtag$ scaled by the sample purity, $\pb$, as
extracted from the data.
This calculation is performed separately
for electrons and muons. By replacing all occurrences of 
$\bl$ by $\bcl$ in Equation \ref{eqn:BR},
one obtains the corresponding equation for $\brcl$. 
% 
%%%%%%%%%%%%%%%%%%%%%%%%%%%%%%%%%%%%%%%%%%%%%%%%%%%%%%
\subsection{Determining the lepton sample composition}
%%%%%%%%%%%%%%%%%%%%%%%%%%%%%%%%%%%%%%%%%%%%%%%%%%%%%%
\label{sec:extfrac}
%           -----------------
\leavevmode\indent
The fractions of $\bl$ and $\bcl$ decays are extracted from the
data by performing a binned log-likelihood fit to the 
data distributions of the neural network outputs using the
shapes for each contribution obtained from the Monte Carlo.
Two-dimensional distributions 
are formed using the neural network outputs $\nnbl$ and $\nnbcl$
with $20$ by $20$ bins. 
There are five fit parameters: 
the Peterson b fragmentation model parameter $\peterson$, 
$\fbe$, $\fbu$, $\fbce$ and $\fbcu$.
The remaining fraction of selected candidates is assigned to the
background sample: $\bcbarl$ leptons, non-prompt leptons and fake leptons in
$\bbbar$ events, and leptons found in charm and light flavour events. Their
fraction is fixed to be $1 - \fbl - \fbcl$ in the fit for electrons
and muons separately.
The Peterson b fragmentation model parameter $\peterson$
is allowed to float in the fit, with a common value used for both
electrons and muons, to greatly reduce the contributions to the
systematic uncertainties from the modelling of this parameter.
This is done by reweighting the Monte Carlo 
events such as to produce the desired fragmentation function.

%           ---------------------
\subsection{Results}  
%           ---------------------
\label{sec:fitresults}
In total, $29 \thinspace 516$ electrons and $44 \thinspace 832$ muons
are selected in the data opposite to a b-tagged hemisphere. The results of the fit are
$\fbe=0.5726 \pm 0.0042$ and $\fbce=0.2596 \pm 0.0055$ for electrons
and $\fbu=0.4620 \pm 0.0034$ and $\fbcu=0.2166 \pm 0.0051$ for muons
where the uncertainties are statistical. The Peterson
fragmentation parameter is determined to be $\peterson = 0.00573 \pm
0.00040 (\mathrm{stat.})$. These results include some small
corrections described in the next section.
Using these corrected parameters,
the fit can adequately reproduce the observed distributions 
of the neural network output variables in data for
all selected leptons, as shown above in Figures \ref{fittedbu} and
\ref{fittedbcu}. 

The correlation coefficient between $\fbl$ and $\fbcl$ is $-0.35$ for
electrons and $-0.26$ for muons. The correlations of $\fbe$ and $\fbu$ 
with the common fragmentation parameter are $0.32$ and $0.39$
respectively, whilst the fragmentation parameter produces a $0.13$
correlation between $\fbe$ and $\fbu$.
The size of all other correlations is below 0.06.

From Equation \ref{eqn:BR} using the b purity $\pb$ as determined in
section \ref{bpurity}, the semileptonic branching fractions are then
determined to be 
\begin{eqnarray*}
\bre  & = & (10.78\pm 0.08)\pc,\\
\bru  & = & (10.96\pm 0.08)\pc,\\
\brce & = & (8.37\pm 0.18)\pc,\\
\brcu & = & (8.17\pm 0.19)\pc,\\
\end{eqnarray*}
where the errors are statistical.

%%%%%%%%%%%%%%%%%%%%%%%%%%%%%%%%%%%%%%%%%%%%%%%%%%%%%%
\section{Systematic and modelling uncertainties} % on the fit fractions}
%%%%%%%%%%%%%%%%%%%%%%%%%%%%%%%%%%%%%%%%%%%%%%%%%%%%%%
\label{sec:fitsyst}
In this section, possible systematic shifts and uncertainties on
the five fit parameters
are estimated by studying detector modelling and experimental
systematic effects on the results.
The corresponding systematic uncertainties on $\brl$, $\brcl$ and
$\xe$ are compiled in Table~\ref{all-systs}.
%           ---------------------
\subsection{Modelling the semileptonic decay lepton spectrum}
\label{sec:modellept}
%           ---------------------
Previous $\brl$ measurements \cite{OPAL,ALEPH,DELPHI,L3} depended
heavily on the proper modelling of the semileptonic
decay spectrum. The exact shape of the lepton momentum spectrum is
not known and little theoretical progress has been made in recent
years. The use of neural networks to separate $\bl$ and $\bcl$ events from
the background reduces the dependence of the branching ratio measurements
on the shape of the lepton spectrum by making use
of extra information such as jet shape and charge correlations.
Nevertheless, the simulation of the b hadron decays and the prediction
of the lepton momentum spectrum is still a large source of
uncertainty.

Different decay models are used to estimate the effects
of the modelling on the fitted fractions. 
The events are reweighted to reproduce the lepton momentum spectrum in
the rest frame of the b hadron as predicted by the different models.
The ACCMM model \cite{ACCMM} predictions are used for $\bl$ and $\bcl$
decays to calculate the central values of $\brl$ and $\brcl$, using the
prescription in \cite{EWWG-NIM}. The ISGW \cite{ISGW} and
ISGW$^{**}$ models\footnote{This corresponds to a modification of the
ISGW model introduced by the CLEO collaboration whereby all P-wave
contributions (the $\dds$ contributions) are increased from the
nominal $11\pc$ derived in the original model to $32\pc$ to better
describe their data.} 
provide the $\pm 1 \sigma$ deviations for the $\bl$ decays \cite{EWWG-NIM}. 
Although these models were derived using only $\Bo$ and $\Bpl$ mesons,
all b hadrons are reweighted. This has a very small effect on the
central value but gives a more conservative estimate of the
uncertainty on the modelling error than when only the $\Bo$ and $\Bpl$
decays are reweighted. 
The agreement between the data and these and other semileptonic decay
models is further investigated in the Appendix.% \ref{newmodels}.

For the cascade decays, $\bcl$, we
use variations based on the CLEO measurements of the $\mathrm{b}
\rightarrow \mathrm{D}$ spectrum combined with the
ACCMM prediction for $\cl$ decays as described in \cite{EWWG-NIM}.
Since the b-tagging requirement highly suppresses contributions
from charm and light flavours, the error from the semileptonic decay
modelling in charm events is negligible, and these are simply
reweighted to the central ACCMM model as described in \cite{EWWG-NIM}. 
Helicity effects in D$^*$ decays are not taken into account but
are expected to be very small. 

Because of the minimum momentum cut of $2 \gevoc$ imposed on the
selected leptons, the measured lepton identification efficiencies correspond
to a restricted momentum range. The effect of the extrapolation below
the minimum momentum cut-off is taken into account when evaluating
the lepton selection efficiency corresponding to the different models.
% 
%           ---------------------
\subsection{Fragmentation models}
\label{sec:fragmod}
%           ---------------------
Several models have been proposed to describe the heavy quark fragmentation
process. The function of Peterson \etal\ \cite{peterson} was used
to simulate fragmentation
in $\bbbar$ and $\ccbar$ events in the Monte Carlo. For b hadrons, the Peterson parameter is
determined from the fit by reweighting the Monte Carlo events.
For c hadrons, the parameter is varied between
0.025 to 0.031 to obtain a $\xe$ for charm hadrons of
$0.484 \pm 0.008$ \cite{EWWG-NIM}. As suggested in \cite{EWWG-NIM}, 
the models of Collins and Spiller \cite{cas}, and
Kartvelishvili \etal\ \cite{kart} are used to estimate
the systematic uncertainties coming from the shape of the
b quark fragmentation function, quoted as the $+1\sigma$ and
$-1\sigma$ errors respectively. These models also have one free
parameter. The Monte Carlo is reweighted to simulate
each function and the free parameter as determined from the fit.
The effects of the differing fragmentation functions and fitted
parameters on the lepton efficiencies are also included in the
fragmentation modelling error.
The systematic uncertainties associated with the b fragmentation
models are determined from the observed variations in the derived
values of branching fractions obtained with the various functions.
The agreement between the data and these functions is further discussed
in the Appendix.
% 
%           ---------------------
\subsection {Lepton detection efficiencies}
%           ---------------------
The electron detection efficiencies in data and Monte
Carlo are compared for two pure electron samples: a low-momentum
electron sample coming from photon conversions, and a high-momentum
electron sample from Bhabha events. 
From these comparisons and from studying the neural network input
variable distributions,
a systematic uncertainty on the electron detection efficiency of
$4\pc$, and an uncertainty on photon conversion
rejection of $0.8\pc$ is obtained \cite{rb98}.

The muon detection efficiency in data and Monte Carlo are compared for
two samples: muon pairs from two-photon interactions and muons from 
Z$^0 \to \mu^+ \mu^-$ decays. 
The first sample yields muons in the momentum
range of $2$ to $6 \gevoc$, while the second sample gives muons with
momentum greater than $30 \gevoc$.
In the data, $57\pc$ of the selected muon sample is found in the lower range
while only $0.2\pc$ has momentum above $30 \gevoc$. 
By comparing the selected muon samples, the Monte Carlo was
found to slightly underestimate the efficiency in the data and so a
multiplicative correction factor of 1.013 is applied to the measured
efficiencies. A systematic uncertainty of $1.9\pc$ is assigned to the
muon detection efficiencies.

The systematic uncertainty associated
with the momentum extrapolation below $2 \gevoc$ is 
discussed in Section \ref{sec:modellept}.
%
%           ---------------------
\subsection{Detector resolution effects}
%           ---------------------
The tracking resolution and reconstruction efficiency could be slightly different
for data and Monte Carlo. The reconstructed track parameters are smeared by
$\pm10\pc$ in the Monte Carlo and the lepton detection efficiencies
and the fit fractions recalculated. The difference from the final
values for $\brl$, $\brcl$ and the
b fragmentation parameter are used as estimate of this source of
systematic uncertainty.
\subsection{Fake lepton rates}
%           ---------------------
To study the fake rate in the muon sample, three different
samples are used: identified pions in $K^0_s \to \pi^+\pi^-$ decays, three
prong $\tau$ decays, and a kaon enriched sample based on $\dedx$ requirements. 
From a comparison of the fake muon rates in data and Monte Carlo for
these samples, it was determined that a correction factor of
$1.11 \pm 0.12$ must be applied to the Monte Carlo events.
Accordingly, the relative fraction of fake muon events
is changed by $+11\pc$ using a reweighting technique.
The weights are varied by $\pm 12\pc$ and the data refitted.
For electrons, studies on the fake rates
were conducted using samples of photon conversion electrons
and Bhabha events \cite{rb98}. The fake rates measured in the data and
Monte Carlo were consistent, such that no correction was required, but
an uncertainty of $\pm 21\pc$ is assigned for the fake electron rate.
\subsection{b tagging purity}
%           ---------------------
The systematic uncertainty on the value of the b purity
obtained from the data is discussed in Section \ref{bpurity}. This
constitutes a $0.49\pc$ relative error on the final values for 
$\brl$ and $\brcl$. The errors on the b purity given in Table
\ref{systbpur} have been further subdivided 
in Table \ref{all-systs} to show separately
the contributions from uncertainties in $\rc$, $\rb$, $\xec$ (the
mean fraction of the beam energy carried by the weakly decaying charmed
hadrons), gluon splitting to $\bbbar$ and $\ccbar$ pairs, the
branching fraction of charmed mesons into $\Ks$, charmed mesons lifetimes,
decay multiplicities of charmed mesons and charm production fractions.
The errors resulting from the uncertainty in the b purity due to
detector resolution and finite Monte Carlo statistics are combined
with the other contributions from these sources of error.
% 
%           ---------------------
\subsection {Other sources of systematic uncertainties}
\label{sec:othersyst}
%           ---------------------
%
Several other sources of systematic uncertainties have been
investigated. The Monte Carlo is reweighted to simulate the 
desired parameters and the fit is repeated to assess the contributions to the
systematic uncertainty. These sources are summarised in Table \ref{all-systs}.
Since their effects on the branching ratios
and $\xe$ are small, they are only described briefly here.

{\bf Finite Monte Carlo sample size}: This includes contributions from
the evaluation of $\ecc$, $\euds$ and $C_b$ from Table \ref{systbpur}.

{\bf b hadron species:} 
The production fraction for $\Lb$ baryons\footnote{$\Lb$ denotes
all weakly-decaying b baryons produced in Z$^0$ decays} is set to
$(10.1 ^{~+3.9}_{~-3.1})\pc$ \cite{pdg98} and the
semileptonic branching fraction for inclusive $\Lb$ 
is taken to be $(7.4\pm 1.1)\pc$ \cite{sllb,aleph_lb}.

{\bf \boldmath \btoul~ \boldmath transitions:}
The $\btoul$ branching fraction is set to $(1.84 \pm 0.79)\times
10^{-3}$, the combined value of \cite{btou}.

{\bf \boldmath $\mathrm{B}^{0} - \overline{\mathrm{B}^{0}}~$ \unboldmath mixing:}
The $\mathrm{B}_{d}^{0} - \overline{\mathrm{B}_{d}^{0}}$
mixing parameter is set to $\chi_d=0.172 \pm 0.010$ \cite{pdg98} whilst $50\%$
$\mathrm{B}_{s}^{0} - \overline{\mathrm{B}_{s}^{0}}$ mixing is assumed \cite{pdg98}. 

{\bf Fake lepton spectrum:}
Small variations in the distributions of fake leptons are allowed by
shifting the momentum spectrum of fake leptons and
non-prompt lepton in
$\bbbar$ events by $25 \mevoc$, roughly $\pm 0.5\pc$ of the mean momentum
for fake and non-prompt leptons in $\bbbar$ events. 

{\bf Contributions from \boldmath $\btaul$:\unboldmath} 
The $\btaul$ branching fractions are set to $(0.463 \pm
0.071)\pc$ and $(0.452 \pm 0.069)\pc$ \cite{pdg98} for
electrons and muons, respectively.

{\bf Contributions from \boldmath $\bcbarl$: \unboldmath}
This branching fraction is set to $(1.62^{~+0.44}_{~-0.36})\pc$ as
derived in \cite{TN557}.

{\bf Contributions from \boldmath $\bjpsil$: \unboldmath}
The experimental value given 
in \cite{pdg98} for $\mathrm{BR} (b \to J/\psi)$ is $(1.16 \pm 0.10)\pc$.
Combined with a recent BES measurement for 
$\mathrm{BR}(J/\psi \to \ell^+\ell^-) = (5.87 \pm 0.10)\pc$
\cite{BES}, this
gives $\mathrm{BR}(\bjpsil) = (0.0681 \pm 0.0060)\pc$. 

{\bf Effect of \boldmath \Lb~\unboldmath polarisation:}
Leptons coming from semileptonic $\Lb$ decays are reweighted to simulate
a spectrum corresponding to $-56\%$ polarisation according 
to \cite{lbpol}. The systematic errors 
are calculated using the polarisation range $-13\%$ to $-87\%$,
the $95\%$ confidence level limits from \cite{lbpol}. 

\begin{table}
\begin{center}
{\small
\begin{tabular}{|cl||c|c||c|c||c|} \hline
\multicolumn{2}{|c||}{Parameter}         & BR$(\be)$      & BR$(\bce)$     & BR$(\bu)$      & BR$(\bcu)$ & $\xe$ \\ \hline\hline
\multicolumn{7}{|c|} {model-dependent sources}\\ \hline
\multicolumn{2}{|c||}{\bl}               & $^{~-0.078}_{~+0.207}$ & $^{~+0.126}_{~-0.211}$ 
                  & $^{~-0.101}_{~+0.221}$ & $^{~+0.206}_{~-0.320}$ & $^{~-0.0051}_{~+0.0081}$ \\
\multicolumn{2}{|c||}{\bcl}              & $^{~-0.072}_{~+0.057}$ & $^{~+0.149}_{~-0.059}$ 
                  & $^{~-0.064}_{~+0.058}$ & $^{~+0.168}_{~-0.048}$ & $^{~+0.0009}_{~-0.0008}$ \\
\multicolumn{2}{|c||}{fragmentation}     & $^{~+0.047}_{~-0.028}$ & $^{~+0.225}_{~-0.144}$  
                  & $^{~+0.096}_{~-0.070}$ & $^{~+0.236}_{~-0.180}$ & $^{~-0.0118}_{~+0.0102}$ \\\hline
\multicolumn{2}{|c||}{total models}      & $^{~+0.220}_{~-0.110}$ & $^{~+0.298}_{~-0.262}$ 
                  & $^{~+0.248}_{~-0.139}$ & $^{~+0.355}_{~-0.370}$ & $^{~+0.0131}_{~-0.0129}$ \\\hline \hline
\multicolumn{7}{|c|} {systematic sources}\\ \hline
\multicolumn{2}{|l||}{lepton efficiencies} & $\mp 0.440$ & $\mp 0.341$ & $\mp 0.208$ & $\mp 0.155$ &  \\
\multicolumn{2}{|l||}{detector}          & $\pm 0.074$ & $\pm 0.113$ & $\pm 0.055$ & $\pm 0.086$ & $\pm 0.0004$ \\
\multicolumn{2}{|l||}{lepton fake rates} & $\pm 0.006$ & $\mp 0.048$ & $\pm 0.037$ & $\mp 0.106$ & $\mp 0.0003$ \\
$\pb$ : &$\rc$     & $\pm 0.022$ & $\pm 0.017$ & $\pm 0.022$ & $\pm 0.017$ &  \\
&$\rb$             & $\mp 0.001$ & $\mp 0.001$ & $\mp 0.001$ & $\mp 0.001$ &  \\
&$\xec$            & $\pm 0.004$ & $\pm 0.003$ & $\pm 0.004$ & $\pm 0.003$ &  \\
&g$\to\bbbar$      & $\pm 0.016$ & $\pm 0.013$ & $\pm 0.016$ & $\pm 0.013$ &  \\
&g$\to\ccbar$      & $\pm 0.010$ & $\pm 0.008$ & $\pm 0.010$ & $\pm 0.008$ &  \\
&BR(D$\to \Ks$)    & $\pm 0.011$ & $\pm 0.008$ & $\pm 0.011$ & $\pm 0.008$ &  \\
&$\Dzero$ lifetime & $\pm 0.002$ & $\pm 0.002$ & $\pm 0.002$ & $\pm 0.002$ &  \\
&$\Dplus$ lifetime & $\pm 0.003$ & $\pm 0.002$ & $\pm 0.003$ & $\pm 0.002$ &  \\
&$\Ds$ lifetime    & $\pm 0.001$ & $\pm 0.001$ & $\pm 0.001$ & $\pm 0.001$ &  \\
&D charged mult.   & $\pm 0.011$ & $\pm 0.008$ & $\pm 0.011$ & $\pm 0.008$ &  \\
&D neutral mult.   & $\mp 0.024$ & $\mp 0.018$ & $\mp 0.024$ & $\mp 0.018$ &  \\
&\fD               & $\pm 0.017$ & $\pm 0.014$ & $\pm 0.017$ & $\pm 0.014$ &  \\
&\fDs              & $\pm 0.002$ & $\pm 0.001$ & $\pm 0.002$ & $\pm 0.001$ &  \\
&\fLc              & $\mp 0.007$ & $\mp 0.005$ & $\mp 0.007$ & $\mp 0.005$ &  \\
\multicolumn{2}{|l||}{MC statistics}     & $\pm 0.019$ & $\pm 0.042$ & $\pm 0.022$ & $\pm 0.049$ & $\pm 0.0010$ \\
\multicolumn{2}{|l||}{b hadron species}  & $\mp 0.013$ & $\pm 0.022$ & $\mp 0.012$ & $\pm 0.030$ & $\mp 0.0006$ \\
\multicolumn{2}{|l||}{\btoul}            & $\pm 0.004$ &             & $\pm 0.009$ & $\pm 0.022$ & $\mp 0.0020$ \\
\multicolumn{2}{|l||}{B mixing}          & $\pm 0.002$ & $\pm 0.016$ & $\mp 0.002$ & $\pm 0.007$ & $\pm 0.0002$ \\
\multicolumn{2}{|l||}{fake lepton spectrum}&             & $\mp 0.003$ & $\mp 0.002$ & $\mp 0.042$ &  \\
\multicolumn{2}{|l||}{\btaul}            & $\mp 0.026$ & $\mp 0.013$ & $\mp 0.021$ & $\mp 0.019$ & $\pm 0.0003$ \\
\multicolumn{2}{|l||}{\bcbarl}           & $\mp 0.004$ & $\mp 0.081$ & $\mp 0.023$ & $\mp 0.064$ & $\pm 0.0003$ \\
\multicolumn{2}{|l||}{\jpsil}            & $\mp 0.004$ &             & $\mp 0.002$ &             & $\pm 0.0001$ \\
\multicolumn{2}{|l||}{\Lb~polarisation}  & $\pm 0.004$ & $\pm 0.006$ & $\pm 0.005$ & $\pm 0.026$ & $^{-0.0013}_{+0.0020}$ \\\hline
\multicolumn{2}{|c||}{experimental systematic}        & $\pm 0.450$ & $\pm 0.377$ & $\pm 0.227$ & $\pm 0.234$ & $^{+0.0031}_{-0.0027}$ \\\hline
\end{tabular}
\parbox{15cm}{\caption {\sl
Summary of all model-dependent and experimental systematic uncertainties on
$\brl$ and $\brcl$ (shown separately for electrons and muons),
and $\xe$. All errors are absolute errors given in percent (except for \xe).
The sign on each contribution indicates the correlation between
this systematic uncertainty and the final results.
\label{all-systs}}}}
\end{center}
\end{table}
% 
%%%%%%%%%%%%%%%%%%%%%%%%%%%%%%%%%%%%%%%%%%%%%%%%%%%%%%
\subsection{Consistency checks}
%%%%%%%%%%%%%%%%%%%%%%%%%%%%%%%%%%%%%%%%%%%%%%%%%%%%%%
To test the fitting procedure,
the Monte Carlo sample is divided into two equal sub-samples.
The first sample is used as a substitute for the real data in the fit 
while the second is retained as the Monte Carlo sample.
The fitted fractions for the first sample can then be compared to the
true information from the Monte Carlo. 
The fitted fractions and the Peterson fragmentation parameter all agree with
the true values within statistics.

Various tests are performed on the data to check the stability of 
the results by varying the selection criteria.
Firstly, the effect of changing the minimum lepton momentum requirement
on the measured value for $\brl$ is investigated.
The minimum lepton momentum cut is increased from the nominal $2.0 \gevoc$
to $5.0 \gevoc$ in steps of $0.5 \gevoc$ and $\brl$ is recalculated each
time. This test was found to yield good agreement when performed using
the Monte Carlo test samples discussed above. 
Similarly, a cut is applied on the neural network $\nnbl$ output variable shown
in Figure \ref{fittedbu}. The cut is
increased by steps of 0.1 from 0.0 to 0.6. 
A similar test is performed to check the stability of the $\brcl$ results
by imposing a cut on the neural network $\nnbcl$ output variable shown
in Figure \ref{fittedbcu}. The cut is increased up to a value of 0.4.
For all these tests, the variations observed are found to be statistically consistent with
the central values calculated for $\brl$ and $\brcl$.

Varying the binning used to perform
the fit has no significant influence on the central values for $\brl$ and $\brcl$.
The central results are derived using $20$ by $20$ bins for
the 2-dimensional fit to $\nnbl$ and $\nnbcl$. This range is varied
from $5$ by $5$ up to $40$ by $40$ bins, yielding consistent values for the
branching fractions.

Lastly, the data are divided into four sub-samples
corresponding to the different years in which the data were taken.
The b-tagging purity is recalculated for each data sub-set separately.
Taking the uncorrelated systematic errors into account, all sub-samples are
found to be statistically consistent with the full data sample.
% 
%%%%%%%%%%%%%%%%%%%%%%%%%%%%%%%%%%%%%%%%%%%%%%%%%%%%%%
\section{Results and conclusions}
\label{mainresults}
%%%%%%%%%%%%%%%%%%%%%%%%%%%%%%%%%%%%%%%%%%%%%%%%%%%%%%
All the relevant quantities needed to calculate the
branching ratios $\brl$ and \newline $\brcl$ are given in
Table \ref{results}. The values
\begin{eqnarray*}
\bre  & = & (10.78\pm 0.08{\mathrm~(stat.)}\pm 0.45{\mathrm~(syst.)}^{~+0.22}_{~-0.11}{\mathrm~(model)})\pc \\
\bru  & = & (10.96\pm 0.08{\mathrm~(stat.)}\pm 0.23{\mathrm~(syst.)}^{~+0.25}_{~-0.14}{\mathrm~(model)})\pc \\
\brce & = & (8.37\pm 0.18{\mathrm~(stat.)}\pm 0.38{\mathrm~(syst.)}^{~+0.30}_{~-0.26}{\mathrm~(model)})\pc\\
\brcu & = & (8.17\pm 0.19{\mathrm~(stat.)}\pm 0.23{\mathrm~(syst.)}^{~+0.36}_{~-0.37}{\mathrm~(model)})\pc
\end{eqnarray*}
are obtained for the semileptonic branching ratios for electrons and
muons, consistent with lepton universality. 
These four branching ratios are combined together to obtain
\begin{eqnarray*}
\brl  & = & (10.83\pm 0.10{\mathrm~(stat.)}\pm 0.20{\mathrm~(syst.)}^{~+0.20}_{~-0.13}{\mathrm~(model)})\pc \\
\brcl & = & (~8.40\pm 0.16{\mathrm~(stat.)}\pm 0.21{\mathrm~(syst.)}^{~+0.33}_{~-0.29}{\mathrm~(model)})\pc,
\end{eqnarray*}
taking into account the full covariance matrix
as in \cite{EWWG-NIM}. The $\brl$ measurement is the most precise
to date at the Z$^0$ resonance whereas $\brcl$ is more
precise than the current world average value of
$(7.8 \pm 0.6)\%$ \cite{pdg98}. The value derived for $\brcl$ 
is outside the range given by $\brce$ and $\brcu$ due to large off-diagonal
terms in the covariance matrix and strong correlations with
the $\brl$ measurement. The statistical error on $\brl$ is larger
than the individual errors on $\bre$ and $\bru$ since the
statistical errors from $\brce$ and $\brcu$ also contribute. 
The full systematic correlation matrix is given in Table \ref{correlations}.

\begin{table}
\begin{center}
\begin{tabular}{|c|c|c|} \hline
         & electrons & muons \\  \hline \hline
\pb      & \multicolumn{2}{|c|} {$0.9190 \pm 0.0002 \mathrm~(stat.) \pm 0.0045 {\mathrm~(syst.)}$} \\ \hline 
\nbtag   & $301303$  & $302577$\\
\nol     & $29516$   & $44832$ \\  \hline\hline
\effl    & $0.5662 \pm 0.0231 {\mathrm~(syst.)}$ 
         & $0.6794 \pm 0.0129 {\mathrm~(syst.)}$ \\
\fbl     & $0.5726 \pm 0.0042 \pm 0.0041 {\mathrm~(syst.)}$  
         & $0.4620 \pm 0.0034 \pm 0.0031 {\mathrm~(syst.)}$   \\\hline
\brl     & $(10.780 \pm 0.079 \pm 0.450 ^{~+0.220}_{~-0.109} )\pc $ 
         & $(10.964 \pm 0.081 \pm 0.227 ^{~+0.248}_{~-0.139} )\pc $ \\ \hline\hline
\effcl   & $0.3306 \pm 0.0135 {\mathrm~(syst.)}$
         & $0.4277 \pm 0.0081 {\mathrm~(syst.)}$\\
\fbcl    & $0.2596 \pm 0.0055 \pm 0.0047 {\mathrm~(syst.)}$ 
         & $0.2166 \pm 0.0051 \pm 0.0045 {\mathrm~(syst.)}$   \\ \hline
\brcl    & $(8.370 \pm 0.177 \pm 0.377 ^{~+0.298}_{~-0.262})\pc $ 
         & $(8.167 \pm 0.192 \pm 0.234 ^{~+0.355}_{~-0.370})\pc $ \\ \hline
\end{tabular}
\parbox{15cm}{\caption {\sl
Results for the data sample including all systematic uncertainties for
electrons and muons. The uncertainties from detector resolution have
been added to the errors on the fitted fractions $\fbl$ and
$\fbcl$. The uncertainties due to semileptonic decay and
fragmentation modelling are shown in the last error on the branching
fractions.
\label{results}}}
\end{center}
\end{table}
\begin{table}
\begin{center}
\begin{tabular}{|l||cccc|}\hline
~~~~~ &  \bre & \brce & \bru    & \brcu \\ \hline\hline
\bre  &   $~~1.00$ &       &         &     \\
\brce &   $~~0.40$ &   $~~1.00$ &         &      \\
\bru  &   $~~0.34$ &  $-0.22$ &  $~~1.00$  &     \\
\brcu &  $-0.26$ &   $~~0.53$ &  $-0.22$  &   $~~1.00$  \\ \hline 
\end{tabular}
\parbox{15cm}{\caption {\sl
The full systematic correlation matrix from the averaging procedure
of $\bre$, $\bru$, $\brce$ and $\brcu$.
\label{correlations}}}
\end{center}
\end{table}

The $\bre$ and $\bru$ measurements presented here are consistent with
the current average of all Z$^0$ measurements, 
${\mathrm BR}_{SL}^{\mathrm b}= (10.99 \pm 0.23)\pc$ \cite{pdg98},
based on a global fit to several
electroweak parameters and including specific measurements of $\brl$
\cite{OPAL,ALEPH,DELPHI,L3}. 
On the other hand, the measurement of $\brl$ is still larger than
the $\upfs$ measurement of 
${\mathrm BR}_{SL}^B = (10.45 \pm 0.21)\pc$ \cite{pdg98},
when it is expected to be lower, as explained in Section
\ref{sec:intro}. If the lifetime ratio correction is applied, the
difference between this result and the $\upfs$ measurement is
about 1.8 standard deviations.

\begin{figure}[htbp]\centering
 \begin{center}
   \mbox{
    \epsfxsize=\textwidth
    \epsffile{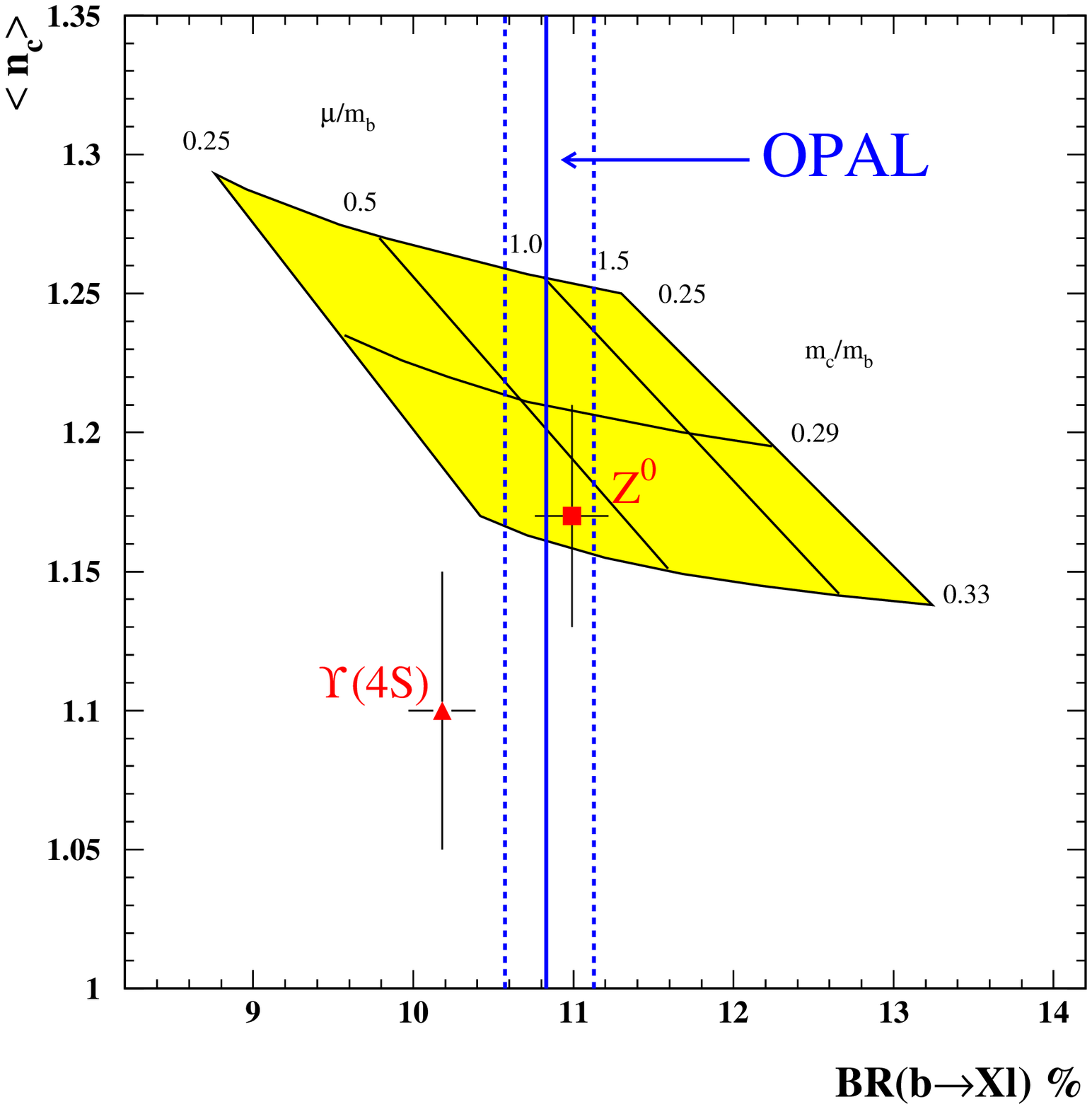} 
   }
 \parbox{15cm}{\caption {\sl
The number of charm hadrons per b hadron decay, $\nc$, as a function
of $\brl$ for the {\rm Z}$^0$ \cite{pdg98} and $\Bell$ for the
$\upfs$ result \cite{pdg98}. The
result derived in this analysis is superimposed as a vertical line
with error bars.
The shaded box represents the theoretical calculation in the framework
of Heavy Quark Effective Theory including higher-order corrections
as given in \cite{neubert} for different assumptions for the ratio
of the renormalisation scale $\mu$ and the b quark mass $m_b$, and
the ratio of the c and b quark masses, $m_c/m_b$. A correction
of $\tau_{\mathrm b}/\tau_{\mathrm B}$ has
been applied to the theoretical prediction and to the $\upfs$
result, as described in the text.
\label{theory}}}
\end{center}
\vspace*{-0.8cm}
\end{figure}

This measurement is also consistent with theoretical calculations as can be seen
in Figure~\ref{theory}, where a correction factor of 0.974
corresponding to the lifetime ratio 
$\tau_{\mathrm b}$/$\tau_{\mathrm B}$ \cite{pdg98} has been applied to both the
theoretical calculations and the $\upfs$ value of $\Bl$ to
allow direct comparison with the Z$^0$ results. No correction is
applied to the values of $\nc$, the average number of charm hadrons
produced per b decay.

From the fitted fragmentation model parameters, the average value of
the fraction of the beam energy carried by the weakly decaying
b hadron is obtained, giving
\begin{eqnarray*}
\xe = 0.709 \pm 0.003 {\mathrm~(stat.)}\pm 0.003 {\mathrm~(syst.)} \pm 0.013 {\mathrm~(model)} 
\end{eqnarray*}
where the modelling error is dominated by the choice of b fragmentation model.

All the measurements presented here are statistically independent of,
and consistent with similar results derived in a previous OPAL
analysis \cite{OPAL} where the quantities $\brl = (10.5\pm
0.6{\mathrm~(stat.)}\pm 0.5{\mathrm~(syst.)})\pc$,  $\brcl
= (7.7 \pm 0.4 \pm 0.7)\pc$ and $\xe = 0.697 \pm 0.006
{\mathrm~(stat.)} \pm 0.011{\mathrm~(syst.)}$ were extracted from a
fit for these and several other parameters (including $\rb$). 
However, the uncertainties related to
assessing the systematic correlations between these old results and
those presented in this paper means that no overall gain in precision
is obtained by combining them.
Therefore the results presented here supersede the
results previously published in \cite{OPAL}.
\par
% \newpage
\section*{Acknowledgements}
\par
We particularly wish to thank the SL Division for the efficient operation
of the LEP accelerator at all energies
 and for their continuing close cooperation with
our experimental group.  We thank our colleagues from CEA, DAPNIA/SPP,
CE-Saclay for their efforts over the years on the time-of-flight and trigger
systems which we continue to use.  In addition to the support staff at our own
institutions we are pleased to acknowledge the  \\
Department of Energy, USA, \\
National Science Foundation, USA, \\
Particle Physics and Astronomy Research Council, UK, \\
Natural Sciences and Engineering Research Council, Canada, \\
Israel Science Foundation, administered by the Israel
Academy of Science and Humanities, \\
Minerva Gesellschaft, \\
Benoziyo Center for High Energy Physics,\\
Japanese Ministry of Education, Science and Culture (the
Monbusho) and a grant under the Monbusho International
Science Research Program,\\
Japanese Society for the Promotion of Science (JSPS),\\
German Israeli Bi-national Science Foundation (GIF), \\
Bundesministerium f\"ur Bildung, Wissenschaft,
Forschung und Technologie, Germany, \\
National Research Council of Canada, \\
Research Corporation, USA,\\
Hungarian Foundation for Scientific Research, OTKA T-029328, 
T023793 and OTKA F-023259.\\
\bigskip
\begin{appendix}
\noindent{\Huge{\bf Appendix}}
%%%%%%%%%%%%%%%%%%%%%%%%%%%%%%%%%%%%%%%%%%%%%%%%%%%%%
\section*{Semileptonic b decay and fragmentation models}
%           --------------------------------------------
For the first time at LEP,
an attempt is made
to probe the level of agreement between the
data and various semileptonic decay models. This test is 
also sensitive enough to allow a closer examination of different
fragmentation functions.
In this section, several theoretical $\bl$ decay models are
investigated. These models affect both the lepton total
and transverse momentum spectra. For each $\bl$ decay model, we use three 
different fragmentation functions, those of Peterson \etal\
\cite{peterson}, Collins and Spiller \cite{cas} and Kartvelishvili
\etal\ \cite{kart}. These functions primarily affect the lepton total
momentum spectrum, leaving the transverse momentum unchanged.
The same models are used to simulate the cascade decays $\bcl$, and to
assess the modelling uncertainties, as described in Section
\ref{sec:modellept}. 

\begin{figure}[htbp]\centering
 \begin{center}
   \mbox{
    \epsfxsize=0.85\textwidth
    \epsffile{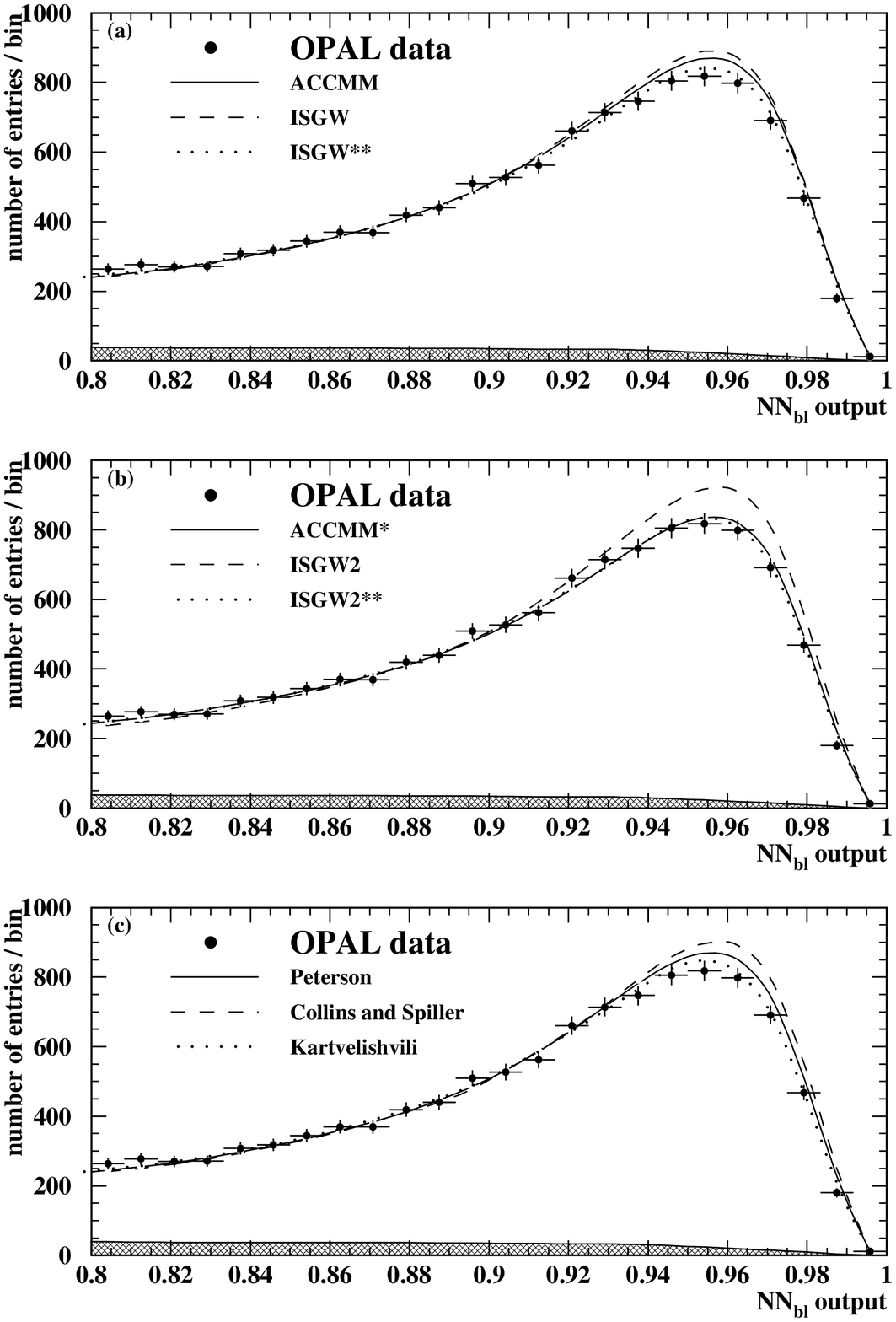}
   }
 \parbox{15cm}{\caption {\sl
The fitted distributions for the $\nnbl$ neural network output for
    electrons with 
(a) the ACCMM, ISGW and ISGW$^{**}$ models;
(b) the ISGW2, ISGW2$^{**}$ and ACCMM$^{*}$ models;
(c) the ACCMM model with the Peterson, Collins and Spiller and
    Kartvelishvili fragmentation functions. 
The Peterson function is used to describe the fragmentation in (a) and (b).
The shaded area shows contributions from sources other than $\be$ in
the data, as extracted from the fit.
\label{fig:modl-e}}}
\end{center}
\vspace*{-0.8cm}
\end{figure}

\begin{figure}[htbp]\centering
 \begin{center}
   \mbox{
    \epsfxsize=0.85\textwidth
    \epsffile{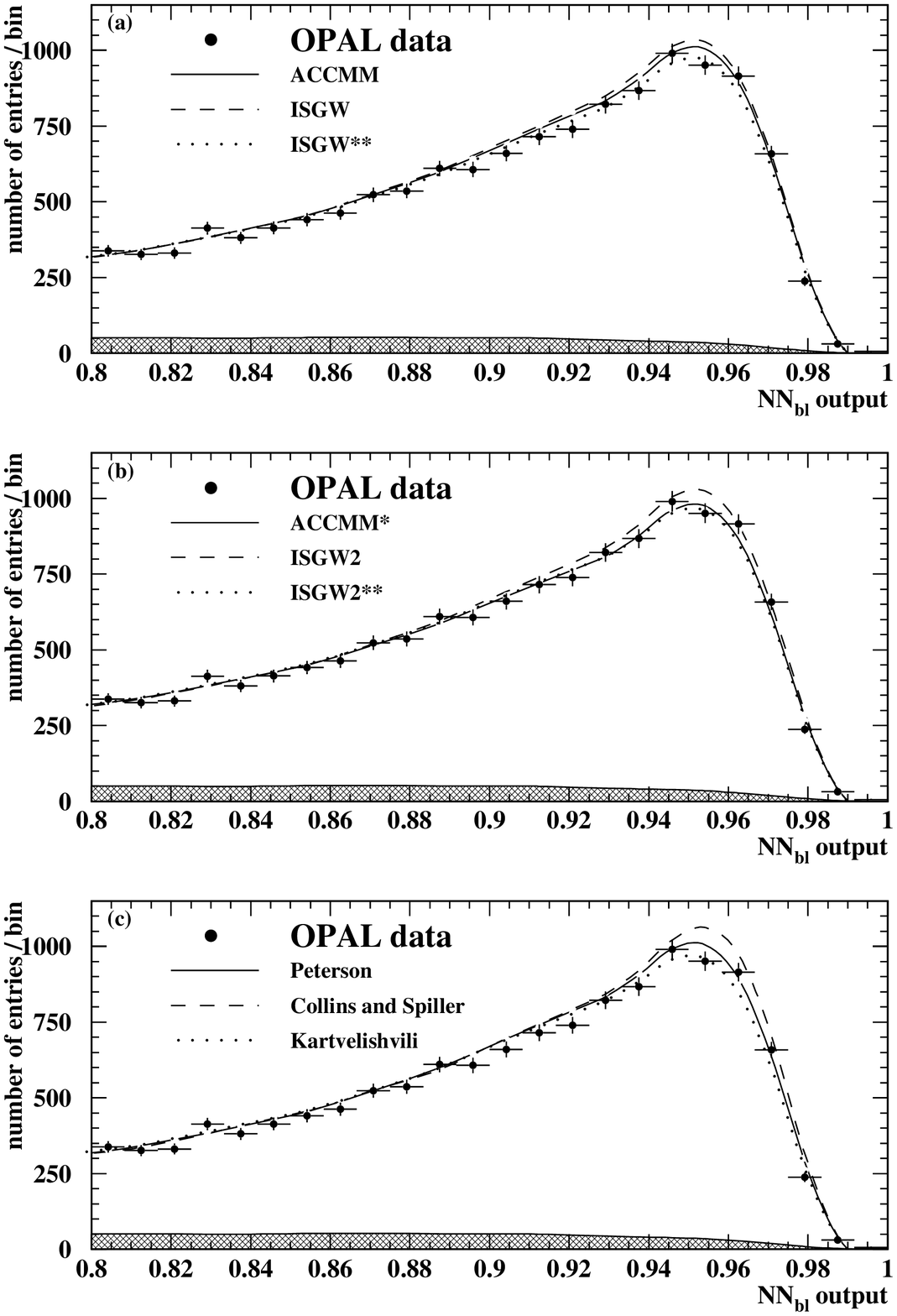}
   }
 \parbox{15cm}{\caption {\sl
The fitted distributions for the $\nnbl$ neural network output for
    muons with 
(a) the ACCMM, ISGW and ISGW$^{**}$ models;
(b) the ISGW2, ISGW2$^{**}$ and ACCMM$^{*}$ models;
(c) the ACCMM model with the Peterson, Collins and Spiller and
    Kartvelishvili fragmentation functions. 
The Peterson function is used to describe the fragmentation in (a) and (b).
The shaded area shows contributions from sources other than $\bu$ in
the data, as extracted from the fit.
\label{fig:modl-m}}}
\end{center}
\vspace*{-0.8cm}
\end{figure}

The six $\bl$ decay models investigated are:
\begin{enumerate}
\item ACCMM model \cite{ACCMM}: all parameters were tuned to the CLEO
  data \cite{cltune}. Their values are
  fixed as given in \cite{EWWG-NIM}: the Fermi 
  momentum of the spectator quark, $\pf = 298 \mevoc$, the mass of the charm
  quark, $\mc = 1673 \mevocc$, and the mass of the spectator quark,
  $\msp =150 \mevocc$.

\item ISGW model \cite{ISGW}: this model has no free parameters and the
  $\dds$ contributions are predicted to account for $11\pc$ of all b
  decays.

\item ISGW$^{**}$ model: the ISGW model modified such as to allow
  the total contributions from $\dds$ to account for $32\pc$ of all b decays, as
  described in \cite{EWWG-NIM}.
   
\item ISGW2 \cite{isgw2}: a revised version of the ISGW model incorporating
  constraints from heavy quark symmetry, hyperfine distortions of
  wave functions, and form factors with more realistic high-recoil
  behaviour. This model 
  predicts that the sum of all $\dds$ contributions accounts for
  $9.3\pc$ of the total b decay width.
  The $\btoul$ branching ratio was fixed to $2.2\pc$ of all b decays and
  mixing was assumed between the $\eta$ and $\eta^{'}$ final states.
\item ISGW2$^{**}$: the ISGW2 model modified to allow the sum
  of all $\dds$
  contributions to be a free parameter of the fit. Best agreement with
  OPAL data is found when the $\dds$ contribution amounts to  
$(45 \pm 3 {\mathrm~(stat.)} \pm 3{\mathrm~(syst.)} \pm 4 \mathrm{~(model)}) \pc$ of
  the total width, instead of the $9.3\pc$ set in the original
  model. The model error contains uncertainties from both the b
  fragmentation (following the same procedure as described in 
  Section \ref{sec:fragmod}) and $\bcl$ decay models.
  The Peterson fragmentation model is used to derive the central value. 
\item ACCMM$^*$: the ACCMM model with free parameters. The Fermi 
  momentum of the spectator quark, $\pf$, and the mass of the charm
  quark, $\mc$, are treated as free parameters in the fit, giving
  $\pf=(837 \pm 143 {\mathrm~(stat.)} \pm 132 {\mathrm~(syst.)}
 ^{~+234}_{~-186} \mathrm{~(model)} )\mevoc$ and
  $\mc=(1287 \pm 100 {\mathrm~(stat.)}\pm 87 {\mathrm~(syst.)}
  ^{~+112}_{~-136}\mathrm{~(model)}) \mevocc$. 
  The mass of the spectator quark is kept fixed at $150 \mevocc$ as in
  \cite{EWWG-NIM}.

  The semileptonic decay model parameters $\pf$ and $\mc$ strongly depend
  on the choice of the fragmentation model used. The Peterson model is
  used to derive the central values. The modelling errors
  given here again correspond to the b fragmentation and
  $\bcl$ decay model errors added in quadrature.
  The correlation coefficient between these model parameters
  is $-0.970$. The decay model parameters are consistent
  with the calculated $\pf$ value of about $550 \mevoc$ derived
  in \cite{pfth} using the relativistic quark model, and the world
  average charm mass of $1100$ to $1400 \mevocc$ taken from
  \cite{pdg98}. The $\btoul$ branching ratio was fixed to $2.7\pc$ of all b decays.
\end{enumerate} 
Figures \ref{fig:modl-e} and \ref{fig:modl-m} show the fitted
distributions for each of these models compared to the data in the
$\bl$ peak region of the $\nnbl$ neural network output distributions
for electrons and muons, respectively.  The fit is performed over the
full range of the neural network output (from zero to one). 
The region shown corresponds to $\nnbl> 0.8$, 
which, from the fit results to the data, is approximately $93\pc$ pure in $\bl$ decays.

The results for $\brl$ and $\brcl$ obtained with these
different models are summarised in Table \ref{bl:models} together with
the statistical, systematic and modelling uncertainties.
All errors are calculated according to the procedures outlined in the
preceding sections, apart from the modelling error which accounts for
$\bcl$ decay modelling only.
The values obtained for the decay model parameters as well as for the
free parameter in the fragmentation functions are also given in Table
\ref{bl:models}. The results for $\xe$ corresponding to the various
fitted fragmentation functions are also listed. These can be compared
to the value of $\xe = 0.702 \pm 0.008$ from \cite{TN557} obtained
from a multi-parameter fit to several electroweak parameters.
The $\chi^2/$bin is calculated using the portion of the $\nnbl$ output
shown in Figures \ref{fig:modl-e} and \ref{fig:modl-m} using 
statistical, systematic and modelling uncertainties from both the
electron and muon samples. These are given as an indicator of the
agreement between these models and the data.

\renewcommand{\arraystretch}{1.5}
\begin{table}
\begin{sideways}
\begin{minipage}[b]{\textheight}
\centering
{\scriptsize
\begin{tabular}{|l|c||c|c||c|c|c|c||c|} \hline
$\textvl{\bl}{model}$  & $\textvl{\bl~model}{parameters}$ & $\textvl{Fragmentation}{parameter}$ & $\xe$ &
$\bre$ & $\bru$  & $\brce$ & $\brcu$ & $\chi^2/$bin \\ \hline\hline
\multicolumn{9}{|c|} {Peterson \etal\ } \\ \hline
ACCMM & fixed &$0.00573 \pm 0.00062$ & $0.709 \pm 0.004$ &$10.78 \pm 0.08 \pm 0.45 ^{~-0.07}_{~+0.06}$& $10.96 \pm 0.08 \pm 0.23 ^{~-0.06}_{~+0.06}$ & $8.37 \pm 0.18 \pm 0.38 ^{~+0.15}_{~-0.06}$  & $8.17 \pm 0.19 \pm 0.23 ^{~+0.17}_{~-0.05}$ & 64/48\\
ISGW  & fixed &$0.00655 \pm 0.00070$ & $0.705 \pm 0.004$ &$10.70 \pm 0.08 \pm 0.45 ^{~-0.07}_{~+0.06}$ & $10.86 \pm 0.08 \pm 0.23 ^{~-0.07}_{~+0.06}$ &$8.50 \pm 0.18 \pm 0.38 ^{~+0.16}_{~-0.06}$  & $8.37 \pm 0.19 \pm 0.24 ^{~+0.16}_{~-0.04}$ & 98/48\\
ISGW$^{**}$ & fixed &$0.00456 \pm 0.00051$ & $0.718 \pm 0.004$ &$10.99 \pm 0.08 \pm 0.46 ^{~-0.07}_{~+0.06}$ & $11.19 \pm 0.08 \pm 0.23 ^{~-0.06}_{~+0.06}$ &$8.16 \pm 0.18 \pm 0.37 ^{~+0.14}_{~-0.06}$  & $7.85 \pm 0.20 \pm 0.23 ^{~+0.19}_{~-0.07}$ & 37/48 \\
ISGW2 & fixed &$0.00787 \pm 0.00083$ & $0.698 \pm 0.004$ &$10.69 \pm 0.08 \pm 0.45 ^{~-0.07}_{~+0.06}$ & $10.86 \pm 0.08 \pm 0.23 ^{~-0.07}_{~+0.06}$ &$8.62 \pm 0.18 \pm 0.38 ^{~+0.16}_{~-0.06}$  & $8.55 \pm 0.19 \pm 0.24 ^{~+0.15}_{~-0.03}$ & 131/48 \\
ISGW2$^{**}$& $f_{\dds} = 45 \pm 5$ &$0.00446 \pm 0.00055$ & $0.719 \pm 0.004$ &$10.95 \pm 0.08 \pm 0.45 ^{~-0.09}_{~+0.07}$ & $11.15 \pm 0.08 \pm 0.23 ^{~-0.08}_{~+0.07}$ &$8.17 \pm 0.18 \pm 0.37 ^{~+0.19}_{~-0.07}$  & $7.85 \pm 0.20 \pm 0.23 ^{~+0.40}_{~-0.08}$ & 35/48 \\
ACCMM$^*$ & $\vnbr{p_F = 837 ^{~+204}_{~-217}}{m_c = 1287 ^{~+142}_{~-135}}$ &$0.00465 \pm 0.00054$ & $0.717 \pm 0.004$ &$10.95 \pm 0.09 \pm 0.46 ^{~-0.11}_{~+0.08}$ & $11.15 \pm 0.09 \pm 0.23 ^{~-0.10}_{~+0.08}$&$8.16 \pm 0.18 \pm 0.37 ^{~+0.23}_{~-0.08}$ & $7.84 \pm 0.20 \pm 0.23 ^{~+0.51}_{~-0.09}$ & 38/48\\ \hline\hline
\multicolumn{9}{|c|} {Collins and Spiller} \\ \hline
ACCMM & fixed &$0.00342 \pm 0.00062$ & $0.698 \pm 0.004$ &$10.83 \pm 0.08 \pm 0.45 ^{~-0.07}_{~+0.06}$ & $11.06 \pm 0.08 \pm 0.23 ^{~-0.07}_{~+0.06}$ &$8.60 \pm 0.18 \pm 0.38 ^{~+0.15}_{~-0.06}$  & $8.40 \pm 0.19 \pm 0.24 ^{~+0.18}_{~-0.06}$ & 148/48\\
ISGW & fixed &$0.00421 \pm 0.00074$ & $0.693 \pm 0.004$ &$10.74 \pm 0.08 \pm 0.45 ^{~-0.07}_{~+0.06}$ & $10.95 \pm 0.08 \pm 0.23 ^{~-0.07}_{~+0.06}$ &$8.72 \pm 0.18 \pm 0.38 ^{~+0.16}_{~-0.06}$  & $8.61 \pm 0.19 \pm 0.24 ^{~+0.17}_{~-0.05}$ & 202/48\\
ISGW$^{**}$ & fixed &$0.00241 \pm 0.00044$ & $0.705 \pm 0.004$ &$11.05 \pm 0.08 \pm 0.46 ^{~-0.07}_{~+0.06}$ & $11.30 \pm 0.08 \pm 0.23 ^{~-0.07}_{~+0.06}$ &$8.39 \pm 0.18 \pm 0.38 ^{~+0.14}_{~-0.06}$  & $8.09 \pm 0.20 \pm 0.23 ^{~+0.20}_{~-0.08}$ & 84/48 \\
ISGW2 & fixed &$0.00556 \pm 0.00096$ & $0.687 \pm 0.004$ &$10.72 \pm 0.08 \pm 0.45 ^{~-0.07}_{~+0.06}$ & $10.94 \pm 0.08 \pm 0.23 ^{~-0.07}_{~+0.06}$ &$8.83 \pm 0.18 \pm 0.39 ^{~+0.16}_{~-0.06}$  & $8.79 \pm 0.19 \pm 0.24 ^{~+0.16}_{~-0.04}$ & 253/48 \\
ISGW2$^{**}$& $f_{\dds} = 43 \pm 5$ &$0.00251 \pm 0.00044$ & $0.704 \pm 0.004$ &$11.00 \pm 0.08 \pm 0.45 ^{~-0.09}_{~+0.07}$ & $11.24 \pm 0.08 \pm 0.23 ^{~-0.08}_{~+0.07}$ &$8.43 \pm 0.18 \pm 0.38 ^{~+0.19}_{~-0.07}$  & $8.14 \pm 0.20 \pm 0.23 ^{~+0.41}_{~-0.06}$ & 80/48 \\
ACCMM$^*$ & $\vnbr{p_F = 679 ^{~+180}_{~-192}}{m_c = 1287 ^{~+146}_{~-138}}$ &$0.00252 \pm 0.00043$ & $0.704 \pm 0.004$ &$10.94 \pm 0.09 \pm 0.45 ^{~-0.11}_{~+0.08}$ & $11.19 \pm 0.09 \pm 0.23 ^{~-0.10}_{~+0.08}$ &$8.43 \pm 0.18 \pm 0.38 ^{~+0.23}_{~-0.08}$  & $8.14 \pm 0.20 \pm 0.23 ^{~+0.49}_{~-0.09}$ & 88/48\\ \hline\hline
\multicolumn{9}{|c|} {Kartvelishvili \etal\ } \\ \hline
ACCMM & fixed &$10.04 \pm 0.57$ & $0.720 \pm 0.005$ &$10.75 \pm 0.08 \pm 0.45 ^{~-0.07}_{~+0.06}$ & $10.89 \pm 0.08 \pm 0.23 ^{~-0.07}_{~+0.06}$ & $8.23 \pm 0.18 \pm 0.37 ^{~+0.15}_{~-0.06}$  & $7.99 \pm 0.19 \pm 0.23 ^{~+0.16}_{~-0.04}$ & 41/48\\
ISGW & fixed &$~9.40 \pm 0.54$ & $0.714 \pm 0.005$ &$10.69 \pm 0.08 \pm 0.45 ^{~-0.07}_{~+0.06}$ & $10.80 \pm 0.08 \pm 0.23 ^{~-0.07}_{~+0.06}$ &$8.36 \pm 0.18 \pm 0.37 ^{~+0.16}_{~-0.06}$  & $8.20 \pm 0.19 \pm 0.24 ^{~+0.15}_{~-0.03}$ & 56/48\\
ISGW$^{**}$ & fixed &$11.23 \pm 0.63$ & $0.729 \pm 0.005$ &$10.94 \pm 0.08 \pm 0.45 ^{~-0.07}_{~+0.06}$ & $11.10 \pm 0.08 \pm 0.23 ^{~-0.06}_{~+0.06}$ &$8.01 \pm 0.18 \pm 0.37 ^{~+0.14}_{~-0.06}$  & $7.66 \pm 0.20 \pm 0.22 ^{~+0.18}_{~-0.06}$ & 48/48 \\
ISGW2 & fixed &$~8.58 \pm 0.49$ & $0.706 \pm 0.005$ &$10.69 \pm 0.08 \pm 0.45 ^{~-0.07}_{~+0.06}$ & $10.81 \pm 0.08 \pm 0.23 ^{~-0.07}_{~+0.06}$ &$8.48 \pm 0.18 \pm 0.38 ^{~+0.16}_{~-0.06}$  & $8.39 \pm 0.19 \pm 0.24 ^{~+0.14}_{~-0.02}$ & 73/48 \\
ISGW2$^{**}$& $f_{\dds} = 46 \pm 5$ &$11.44 \pm 0.67$ & $0.731 \pm 0.005$ &$10.91 \pm 0.08 \pm 0.45 ^{~-0.08}_{~+0.07}$ & $11.07 \pm 0.08 \pm 0.23 ^{~-0.07}_{~+0.06}$ &$8.01 \pm 0.18 \pm 0.37 ^{~+0.19}_{~-0.07}$  & $7.65 \pm 0.20 \pm 0.22 ^{~+0.39}_{~-0.09}$ & 53/48 \\
ACCMM$^*$ & $\vnbr{p_F = 1063 ^{~+409}_{~-368}}{m_c = 1153 ^{~+184}_{~-199}}$ &$10.96 \pm 0.64$ & $0.727 \pm 0.005$ &$10.95 \pm 0.09 \pm 0.46 ^{~-0.11}_{~+0.11}$ & $11.11 \pm 0.10 \pm 0.23 ^{~-0.10}_{~+0.11}$ &$8.00 \pm 0.18 \pm 0.37 ^{~+0.23}_{~-0.10}$  & $7.64 \pm 0.20 \pm 0.22 ^{~+0.52}_{~-0.20}$ & 39/48\\ \hline
\end{tabular}}
\parbox{0.9\textwidth}{\caption {\sl \small
Branching fractions $\brl$ and $\brcl$ (given in $\%$) derived by
comparing the data to three fragmentation functions and various
semileptonic decay models for $\bl$
decays. The uppermost line corresponds to the central results, as
discussed in section \ref{mainresults}.
The quoted errors on the branching ratios correspond to the
statistical, systematic and $\bcl$ modelling errors, respectively. 
The fitted $\bl$ decay model parameters are also given when
appropriate. The fitted b fragmentation function parameters, and the
corresponding values for $\xe$ are presented.
The combined statistical, systematic and $\bcl$
modelling errors are given for all fitted model parameters.
The $\chi^2$/bin is calculated using the portion of the $\nnbl$ output
shown in Figures \ref{fig:modl-e} and \ref{fig:modl-m}, using all
uncertainties from both the electron and muon
samples; these are given as an indicator of the goodness-of-fit. All
models and their parameters are described in the text 
\label{bl:models}}}
\end{minipage}
\end{sideways}
\end{table}
\renewcommand{\arraystretch} {1.2}

% \subsection{Conclusions}
The accuracy of the test does not allow ruling out
specific $\bl$ models, although some trends are clear:
\bi
\item The fragmentation functions of Peterson \etal\ and
Kartvelishvili \etal\ provide equally good fits to the data. The
fragmentation function of Collins and Spiller is generally disfavoured
by our data. 
\item The semileptonic decay models ISGW$^{**}$ and ISGW2$^{**}$ best agree
with our data when used with the b fragmentation models of Peterson
\etal\ or Kartvelishvili \etal\ However, these models are less
theoretically sound since modifications to the original models were
needed to allow the overall fraction of $\dds$ contributions to be a
free fit parameter when this fraction is in fact one of the
predictions of the models.
\item For the ACCMM$^{*}$ semileptonic decay model, the best fit to
the data is obtained with
\bc 
 $\pf=(837 \pm 143 {\mathrm~(stat.)} \pm 132 {\mathrm~(syst.)} 
^{~+234}_{~-186} \mathrm{~(model)} )\mevoc$,
  $\mc=(1287 \pm 100 {\mathrm~(stat.)}\pm 87 {\mathrm~(syst.)}
  ^{~+112}_{~-136}\mathrm{~(model)}) \mevocc$ 
\ec
when the mass of the spectator quark is kept fixed at $150
\mevocc$. These results are derived using the b fragmentation model of
Peterson \etal
\item The ISGW2 model gives a worse agreement with our data than the
ISGW model, with all fragmentation models.
\ei
\end{appendix}
\bigskip
%%%%%%%%%%%%%%%%%%%%%%%%%%%


\begin{thebibliography}{99}
%%%%%%%%%%%%%%%%%%%%%%%%%%%
\bibitem{ball}
        E. Bagan, P. Ball, V.M. Braun, and P. Gosdzinsky,
        \Journal{\NPB}{432}{1994}{3} and \\ \Journal{\PLB}{342}{1995}{362}; \\
        E. Bagan, P. Ball, B. Fiol, and P. Gosdzinsky, \Journal{\PLB}{351}{1995}{546}.
 
\bibitem{neubert}
        M. Neubert, C.T. Sachrajda, \Journal{\NPB}{483}{1997}{339}.

\bibitem{pdg98}
        Review of Particle Physics, C. Caso \etal\ (Particle Data Group),
        \Journal{\EPC}{3}{1998}{1}.

\bibitem{sllb}
        OPAL Collab., R. Akers \etal, \Journal{\ZPC}{74}{1997}{423}. 

\bibitem{bogus}
        Production and decay of b-flavoured hadrons, K. Honscheid,
        \Journal{\EPC}{3}{1998}{522}.

\bibitem{OPAL}
%       CERN-PPE/93-106 (1993)
        OPAL Collab., R. Akers \etal, \Journal{\ZPC}{60}{1993}{199}.
% 
\bibitem{ALEPH}
%       CERN-PPE/94-017 (1994)
%       CERN-PPE/94-017 (1994)
        ALEPH Collab., D. Buskulic \etal, \Journal{\PLB}{384}{1996}{414};\\
        ALEPH Collab., D. Buskulic \etal, \Journal{\ZPC}{62}{1994}{179}.
% 
\bibitem{DELPHI}
%       CERN-PPE/96-49 (1996)
        DELPHI Collab., P. Abreu \etal, \Journal{\ZPC}{66}{1995}{323}.
\bibitem{L3}
%       CERN-PPE/96-49 (1996)
        L3 Collab., M. Acciarri \etal, \Journal{\ZPC}{71}{1996}{379};\\ 
        L3 Collab., M. Acciarri \etal, \Journal{\PLB}{335}{1994}{542};\\
        L3 Collab., O. Adriani \etal, \Journal{\PLB}{292}{1992}{454}. 

\bibitem{detector}
        OPAL Collab., K.\ts Ahmet \etal, \Journal{\NIMA}{305}{1991}{275};\\
        P.P.\ts Allport \etal, \Journal{\NIMA}{324}{1993}{34};\\
        P.P.\ts Allport \etal, \Journal{\NIMA}{346}{1994}{479}.

\bibitem{dedx}
%         O.\ts Biebel \etal, \Journal{\NIMA}{323}{1992}{169};\\
%         M.\ts Hauschild \etal, \Journal{\NIMA}{314}{1992}{74}.
%         Progress in dE/dx techniques used for particle identification,
        M. Hauschild, \Journal{\NIMA}{379}{1996}{436}; 
%         see also  {\tt http://opalinfo.cern.ch/opal/doc/confrep/html/cr250.html}
% 
\bibitem{jetset}
        T. Sj\"{o}strand, Comp.~Phys.~Comm.~{\bf 82} (1994) 74.

\bibitem{peterson}
        C. Peterson, D. Schlatter, I. Schmitt and P.M. Zerwas,
        \Journal{\PRD}{27}{1983}{105}. \\
        The OPAL parametrisation for JETSET 7.4 is given in\\
        OPAL Collab., G. Alexander \etal, \Journal{\ZPC}{69}{1996}{543}.

\bibitem{ACCMM}
        G. Altarelli \etal, \Journal{\NPB}{208}{1982}{365}.

\bibitem{cltune}
        CLEO Collab., S. Henderson \etal, \Journal{\PRD}{45}{1992}{2212}.

\bibitem{EWWG-NIM}
        The LEP Collabs., 
        ALEPH, DELPHI, L3 and
        OPAL, \Journal{\NIMA}{378}{1996}{101}.%, CERN-PPE/96-183 (1997).% 

\bibitem{opalmc}
        OPAL Collab., J. Alexander \etal, \Journal{\ZPC}{69}{1996}{543}.

\bibitem{gopal}
        J.\ts Allison \etal, \Journal{\NIMA}{317}{1992}{47}.

\bibitem{opalmh}
        OPAL Collab., R. Akers \etal, \Journal{\ZPC}{65}{1995}{17}.

\bibitem{rb98}  %PR 257 1998 R_b paper
        OPAL Collab., G. Abbiendi \etal, \Journal{\EPC}{8}{1999}{217}.
%         Eur. Phys. J. C8 (1999) 217-239 

\bibitem{cone} %pr097 CERN-PPE/94-51 (17th March 1994)
        OPAL Collab., R. Akers \etal, \Journal{\ZPC}{63}{1994}{197}.
% 
\bibitem{muid}
%        PR076
        OPAL Collab., P. Acton \etal, \Journal{\ZPC}{58}{1993}{523}.

\bibitem{subjet}
        OPAL Collab., R. Akers \etal, \Journal{\ZPC}{66}{1995}{555}.
% 
\bibitem{OPAL-sumpt}
%       CERN-PPE/93-106 (1993)
        OPAL Collab., R. Akers \etal, \Journal{\ZPC}{61}{1994}{209}.

\bibitem{ISGW}
        N. Isgur, D. Scora, B. Grinstein and M. Wise, \Journal{\PRD}{39}{1989}{799}. 

\bibitem{cas}
        P. Collins and T. Spiller, J. Phys. {\bf G 11} (1985) 1289. 

\bibitem{kart}
        V.G. Kartvelishvili, A.K. Likhoded and V.A. Petrov,
        \Journal{\PLB}{78}{1978}{615}.

\bibitem{aleph_lb}
         ALEPH Collab., R. Barate \etal, \Journal{\EPC}{5}{1998}{205}.
         
\bibitem{btou}
        L3 Collab., M. Acciarri \etal, \Journal{\PLB}{436}{1998}{174}. \\
        ALEPH Collab., R. Barate \etal, \Journal{\EPC}{6}{1999}{555}.

\bibitem{TN557}
        Input Parameters for the LEP Electroweak Heavy Flavour
        Results for Summer 1998 Conferences,
        LEPHF 98-01 (see {\tt http://www.cern.ch/LEPEWWG/heavy/}) used
        for the Combination of Preliminary Electroweak Measurements
        and Constraints on the Standard Model, ALEPH, DELPHI, L3 and
        OPAL collaborations, the LEP Electroweak Working Group, and
        the SLD Heavy Flavour and Electroweak Groups,  CERN-EP/99-015.

\bibitem{BES}
        BES Collab., J.Z. Bai \etal, \Journal{\PRD}{58}{1998}{92}. 

\bibitem{lbpol}
        OPAL Collab., G. Abbiendi et al. 
        \Journal{\PLB}{444}{1998}{539}.

\bibitem{isgw2} D. Scora and N. Isgur, \Journal{\PRD}{52}{1995}{2783}.

\bibitem{pfth} D.S. Hwang, C.S. Kim and W. Namgung,
        \Journal{\PRD}{54}{1996}{5620}.

\end{thebibliography}
\end{document}